\documentclass[%
 reprint,
superscriptaddress,
 amsmath,amssymb,
 aps,
floatfix,
]{revtex4-2}

\usepackage{dcolumn}
\usepackage{bm}
\usepackage{graphicx}
\usepackage{caption}
\usepackage{subcaption}
\usepackage{amsmath}
\usepackage{placeins}
\usepackage{xcolor}
\usepackage{xfrac}
\usepackage{float}
\usepackage{lineno}
\usepackage{mathtools}

\usepackage{comment}

\usepackage{siunitx}

\usepackage{hyperref}

\usepackage[utf8]{inputenc}
\usepackage[T1]{fontenc}
\usepackage{mathptmx}
\usepackage{etoolbox}
\usepackage{mathrsfs}
\usepackage{amsfonts}
\usepackage{booktabs} 
\usepackage{caption}  
\usepackage{threeparttable} 
\usepackage{algorithm}
\usepackage{algorithmicx}
\usepackage{algpseudocode}
\usepackage{listings}
\usepackage{enumitem}
\usepackage{chngcntr}
\usepackage{booktabs}
\usepackage{lipsum}
\usepackage[T1]{fontenc}    
\usepackage{csquotes}       
\usepackage{diagbox}
\usepackage{footnote}

\renewcommand{\selectlanguage}[1]{}

\begin{document}

\preprint{APS/123-QED}

\title{Data-driven turbulent heat flux modeling with inputs of multiple fidelity}

\author{Matilde Fiore}
\email{matilde.fiore@vki.ac.be}
\affiliation{Environmental and Applied Fluid Dynamics Department, von Karman Institute for Fluid Dynamics, Belgium}

\author{Enrico Saccaggi}
\affiliation{Department of Mechanical and Aerospace Engineering, Politecnico di Torino, Italy}
\affiliation{Environmental and Applied Fluid Dynamics Department, von Karman Institute for Fluid Dynamics, Belgium}

\author{Lilla Koloszar}
\affiliation{Environmental and Applied Fluid Dynamics Department, von Karman Institute for Fluid Dynamics, Belgium}

\author{Yann Bartosiewicz}
\affiliation{Institute of Mechanics, Materials and Civil Engineering (IMMC), Universite catholique
de Louvain (UCLouvain), Place du Levant 2, 1348 Louvain-la-Neuve, Belgium}

\author{Miguel A. Mendez}
\affiliation{Environmental and Applied Fluid Dynamics Department, von Karman Institute for Fluid Dynamics, Belgium}

\date{\today}

\begin{abstract}
Data-driven RANS modeling is emerging as a promising methodology to exploit the information provided by high-fidelity data. However, its widespread application is limited by challenges in generalization and robustness to inconsistencies between input data of varying fidelity levels. This is especially true for thermal turbulent closures, which inherently depend on momentum statistics provided by low or high fidelity turbulence momentum models. This work investigates the impact of momentum modeling inconsistencies on a data-driven thermal closure trained with a dataset with multiple fidelity (DNS and RANS).
The analysis of the model inputs shows that the two fidelity levels correspond to separate regions in the input space. It is here shown that such separation can be exploited by a training with heterogeneous data, allowing the model to detect the level of fidelity in its inputs and adjust its prediction accordingly. In particular, a sensitivity analysis and verification shows that such a model can leverage the data inconsistencies to increase its robustness. Finally, the verification with a CFD simulation shows the potential of this multi-fidelity training approach for flows in which momentum statistics provided by traditional models are affected by model uncertainties.  
\end{abstract}

\maketitle


\section{Introduction}

Data-driven methods are progressively entering the field of turbulence modeling as one of the most promising avenues to overcome the modeling barrier reached by traditional approaches in the last decade. A variety of data-driven approaches for the modeling of Reynolds stresses have been recently proposed \cite{ling2016reynolds, jiang2021interpretable, yin2022iterative, mcconkey2022generalizability, geneva2019quantifying}, among which artificial neural networks \cite{geneva2019quantifying, berrone2022invariances, cai2024revisiting, ling2016reynolds, sotgiu2019towards} are widely adopted to handle large databases and generate complex, non-linear input/output mappings. Despite the merits of these approaches, significant limitations still restrict their use for a wide range of Computational Fluid Dynamics (CFD) simulations.

\citet{zhang2023review} highlights the most important of these limitations, including the lack of generality in data-driven closures, i.e., their limited applicability to flows that deviate from the conditions covered by the training. The derived closures usually target specific classes of flows \cite{srivastava2021generalizable,milani2020machine,zhao2020rans} and existence of general or universal data-driven turbulence models is currently being questioned by the community \cite{girimaji2023turbulence} due to the complex dependence of coherent structures on geometry, Reynolds number, and many other local/global conditions. Moreover, the literature points out limitations when data-driven closures are trained offline \cite{mandler2022frozen}, i.e., with high-fidelity data only in a \textit{frozen} mode. This training strategy neglects the influence of the numerical methods and additional sub-models involved in the CFD setup. The frozen training mode causes model-data inconsistencies \cite{mandler2022frozen}, where the quantities taken as input by the model differ during training and testing phases. This limits the accuracy when the model is applied a posteriori. A typical example of the consistency problem is the turbulent time scale $k/\epsilon$ \cite{schmelzer2020discovery}, which is systematically different in high-fidelity and modeled (RANS) data, leading to inconsistencies in the data-driven closure if this quantity is one of the model inputs.

Current research proposes mitigating this issue by partially replacing the critical high-fidelity turbulent statistics with their RANS counterparts. This approach is followed, for example, by \citet{schmelzer2020discovery}, who developed a $k$-corrective frozen training strategy where the turbulent dissipation rate $\varepsilon$ is computed by solving its RANS transport equation, taking all the other statistics from the high-fidelity database and adding a corrective term accounting for errors in the $k$ production term. \citet{weatheritt2017development} proposed a similar approach in which $\omega=\varepsilon/k$ is computed from RANS equations evaluated with high-fidelity mean flow and Reynolds stresses. \citet{yin2022iterative} included the inconsistency between RANS and DNS estimates of $k$ in the outputs of the data-driven model, and used this to inform the data-driven model when it operates in conjuction with a RANS solver.   

In general, addressing model inconsistencies in turbulence modeling requires balancing the physical consistency of the data-driven model against its applicability in standard RANS solvers. On one hand, using modeled statistics enhances the applicability of the closures but inevitably introduces structural errors in the mapping to compensate for input inaccuracies. On the other hand, utilizing high-fidelity data promotes physical consistency and leverages the real correlation among the statistics, but this comes at the expense of applicability when implemented in RANS solvers.

Moving to non-isothermal problems, the modeling of turbulent heat fluxes depends on the momentum turbulence modeling, and the Reynolds stress tensor is inevitably one of the critical inputs for turbulent heat flux models. Consequently, a thermal turbulence model trained with Reynolds stresses derived from high-fidelity data makes significant errors when deployed in a RANS solver, which uses low-fidelity modeling of Reynolds stresses along with the Boussinesq approximation. 

\begin{figure*}[thb!] 
	\centering
	\includegraphics[scale=0.52]{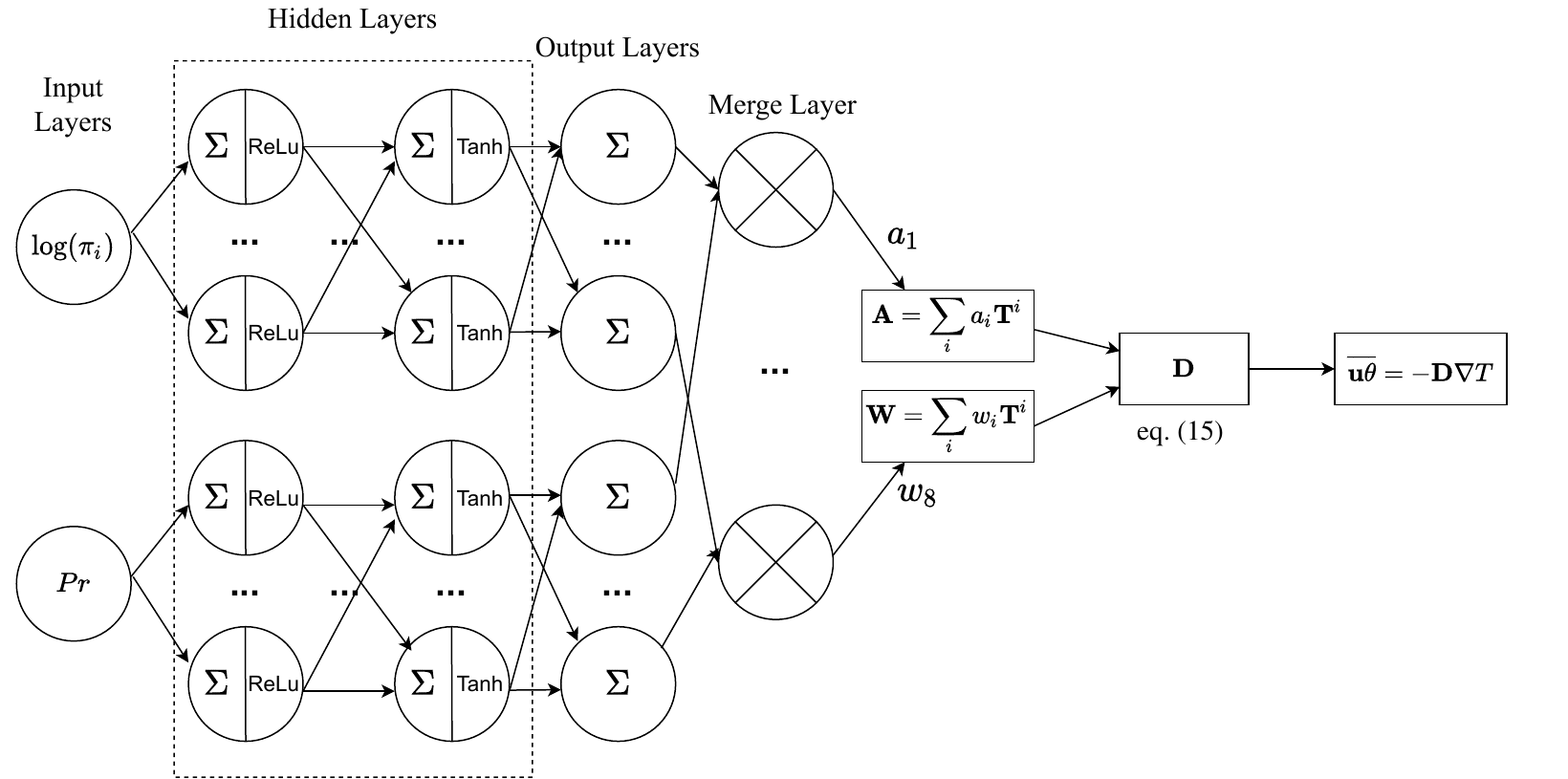}
	\caption{Structure of the artificial neural network used to predict the turbulent heat flux as a function of the molecular Prandtl number and the basis of invariants indicated in Table \ref{tensor_basis}.}
	\label{f:structure}
\end{figure*}

This problem was highlighted in \cite{fiore2022physics}, focusing on the data-driven modeling of the turbulent heat flux at near unity and low Prandtl numbers. Model inconsistencies in the Reynolds stresses were identified to be the major challenge in data-driven thermal closures, partially addressable only by the use of second order momentum turbulence closures. Modeling the Reynolds stresses with the eddy viscosity concept poses significant limitations to thermal turbulence modeling, because the turbulence anisotropy is essential to represent the heat flux vector, and is indeed the foundation of most algebraic thermal closures \cite{daly1970transport,suga2000nonlinear,wikstrom2000derivation}. Such modelling choices are typical of both data-driven and theory-driven RANS modeling. In fact, traditional thermal models are often developed to be combined with specific momentum closures \cite{manservisi2014cfd,manservisi2015cfd}, or variants are proposed to account for different momentum models \cite{shams2019towards}. 

This work explores the hypothesis that machine learning tools could identify the source of the input (e.g. high fidelity vs low fidelity momentum modeling) and adapt accordingly, or find compromises by mitigating its sensitivities to critical inputs. Both approaches could ensure an optimal compromise between physical consistency of the model when this is deployed on high-fidelity data, and robustness when this is deployed with traditional RANS solvers. 

To explore this hypothesis, this paper proposes a training method based on inputs of multiple fidelity multi-objective optimization.
More specifically, the training of a data-driven model for the turbulent heat flux, introduced in section \ref{data-AHFM}, was carried out with a \textit{hybrid} dataset consisting of high-fidelity (DNS) and low fidelity (RANS) data. The peculiarities of such heterogeneous dataset are discussed in section \ref{sec2}. The learning performances are analyzed in section \ref{sec4}: from the model choices within the input space, to the verification of the learned closure and its sensitivity analysis.
The validation of the model is primarily aimed at estimating the robustness gain obtained with this strategy. Based on the results presented in section \ref{results}, conclusions and future development are presented in section \ref{sec6}.

\section{Data-driven AHFM}\label{data-AHFM}

The data-driven Algebraic Heat Flux Model (AHFM) considered in this work is a physics-constrained neural network that predicts the turbulent heat flux at near unity and low Prandtl numbers. Details of its architecture are provided in \cite{fiore2022physics,fiore2022turbulent} and briefly recalled in this section for completeness.

The network embeds rotational invariance properties by construction, as proposed by \citet{ling2016reynolds}, and its predictions are realizable, i.e. they satisfy the second law of thermodynamics. Figure \ref{f:structure} provides a schematic of the neural network structure and its layers. The network predicts variable closure coefficients $a_i$ and $w_i$ based on a set of dimensionless invariants, denoted as $\pi_i$ and defined on the left column of Table \ref{tensor_basis}. Among the quantities listed in the table, $k$ represents the turbulent kinetic energy, $\epsilon$ the turbulent dissipation rate, $k_{\theta}$ the variance of the thermal fluctuations, $\epsilon_{\theta}$ the thermal turbulent dissipation rate. The quantities $\nu$ and $\alpha_l$ denote the molecular viscosity and diffusivity, respectively, while the tensors $\mathbf{S}$, $\boldsymbol{\Omega}$ and $\mathbf{b}$ are defined as:
	\begin{equation}
		\mathbf{S} =\frac{1}{2}\left(\frac{\partial \mathbf{U}}{\partial \mathbf{x}}+\frac{\partial \mathbf{U}}{\partial \mathbf{x}}^T\right),
		\label{S_def}
	\end{equation}

	\begin{equation}
		\boldsymbol{\Omega}=\frac{1}{2}\left(\frac{\partial \mathbf{U}}{\partial \mathbf{x}}-\frac{\partial \mathbf{U}}{\partial \mathbf{x}}^T\right) ,
		\label{Omega_def}
	\end{equation}
\begin{equation}
	\mathbf{b}=\frac{\overline{\mathbf{uu}}}{k}-\frac{2}{3}\mathbf{I},
		\label{b_def}
\end{equation}
in which $\mathbf{U}$ is the mean velocity field, $\mathbf{u}$ the velocity fluctuation and the operator $\overline{\cdot}$ denotes the Reynolds averaging. 

The coefficients $a_i$ and $w_i$ are used to compute the following expansions of tensors:
\begin{align}
    \mathbf{A} &=\sum_{i=1}^n a_i \mathbf{T}^i, & \mathbf{W} &= \sum_{i=1}^n w_i \mathbf{T}^i,
        \label{expansions}
\end{align}
in which $\mathbf{T}^i$ are the element of the tensor basis indicated in Table \ref{tensor_basis}. The dispersion tensor is then computed as a sum of a symmetric, positive-definite tensor and a skew-symmetric tensor:
\begin{equation}
    \mathbf{D}=\left[(\mathbf{A}+\mathbf{A}^T) (\mathbf{A}^T+\mathbf{A}) + \frac{k}{\epsilon^{0.5}} (\mathbf{W}-\mathbf{W}^T)\right],
    \label{chol}
\end{equation}

In Ref. \cite{fiore2022physics,fiore2022turbulent}, the network was trained with high-fidelity DNS data for Prandtl numbers ranging from 0.71 to 0.01. The loss function applied for the training promotes the smoothness of the predicted heat flux fields:
\begin{linenomath}
\begin{equation}
\begin{split}
    \mathcal{L}(\mathbf{q})= \frac{1}{N}\left(\sum_{i=1}^N \sum_{j=1}^3 \Bigl(\hat{q}_{i,j} -q_{i,j}\Bigr)^2 \right)   \\
    + \frac{\lambda}{N} \left( \sum_{i=1}^N \sum_{j,k=1}^3  \left| \frac{\partial \hat{q}_{i,j}}{\partial x_k} - \frac{\partial q_{i,j}}{\partial x_k} \right| \Delta x_k \right),
    \end{split}
    \label{e:of}
\end{equation}
\end{linenomath}
where $i\in[1,\dots N]$ is the index spanning across the $N$ data points contained in each mini-batch, $q_{i,j}=\overline{u_j \theta}$ is the prediction of the ANN and  $\hat{q}_{i,j}$ is the corresponding flux provided by DNS data. Both values $q_{i,j}$ and  $\hat{q}_{i,j}$ are normalized with respect to the maximum DNS value achieved in each flow configuration.

\begin{figure*}[h!]
	\centering
	\subfloat[$\overline{u \theta}$]{%
		\label{tr13}
		\includegraphics[scale=0.36]{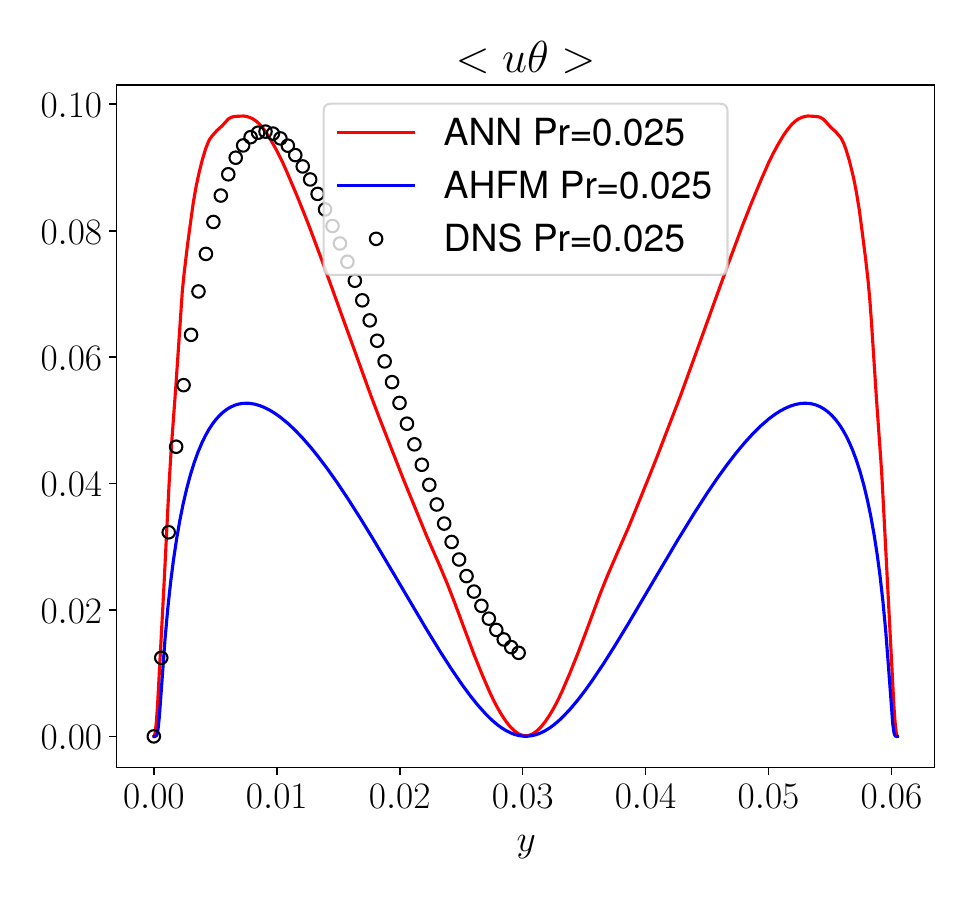}
		\hspace{0.04\linewidth}
		\includegraphics[scale=0.36]{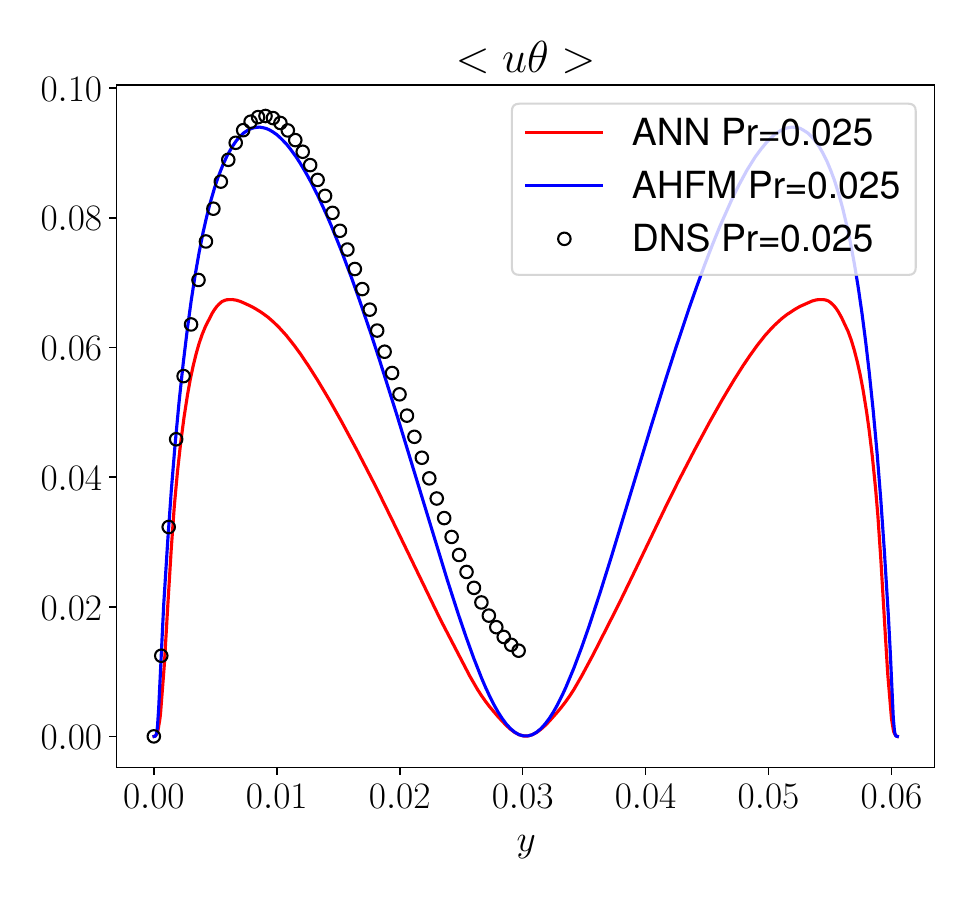}} \\
	\subfloat[$\overline{v \theta}$]{%
		\label{tr14}
		\includegraphics[scale=0.36]{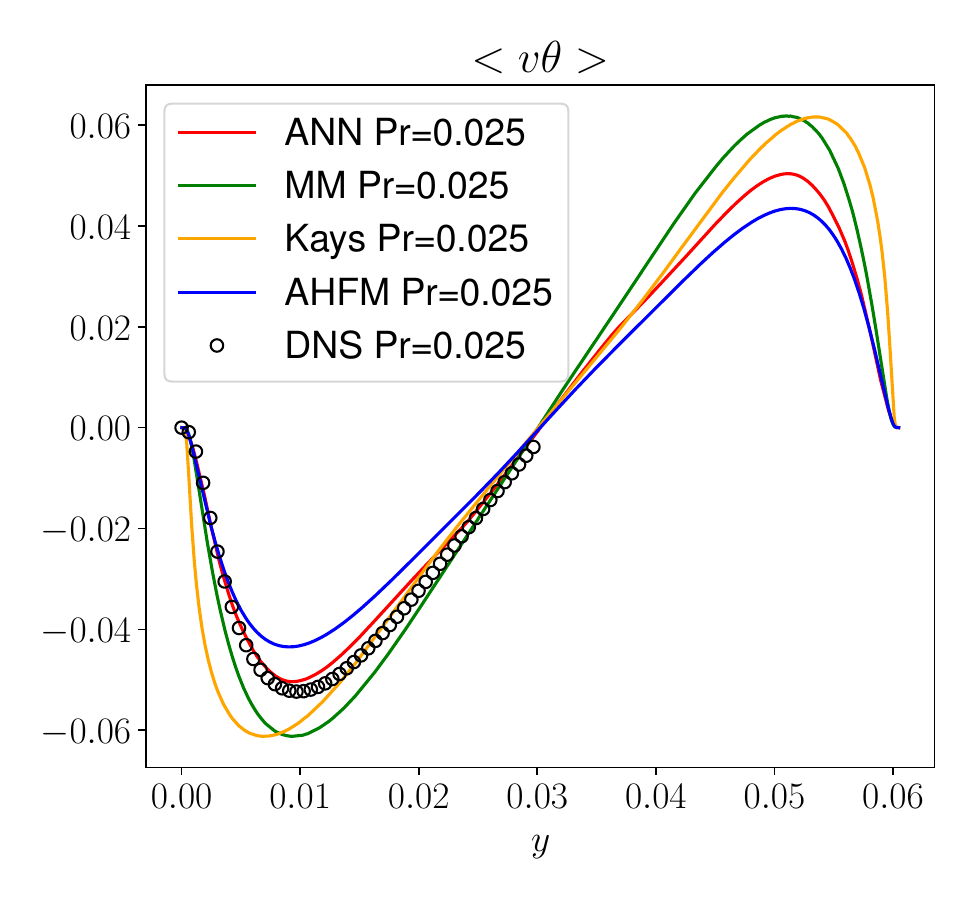}
		\hspace{0.04\linewidth}      
		\includegraphics[scale=0.36]{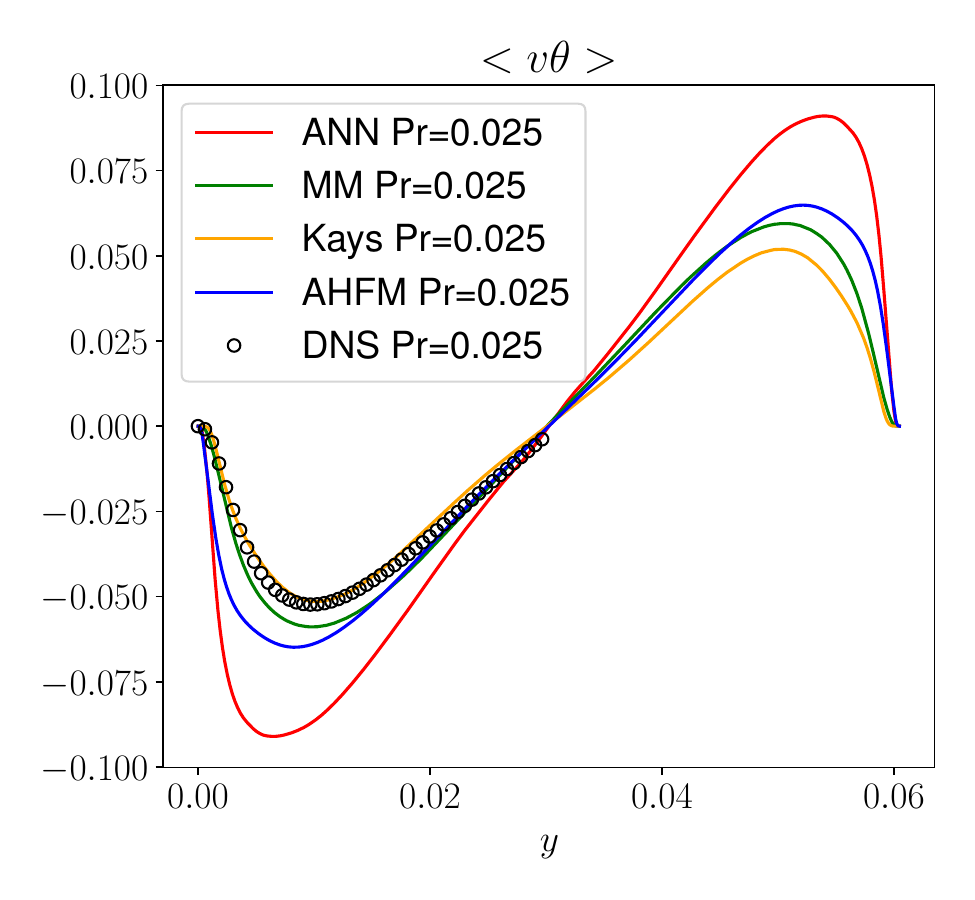}}
	
	\caption{Comparison of the results achieved with the data-driven model (ML), the Manservisi model (MM), the Kays correlation, and the AHFM model at $Re_{\tau}$=395 and $Pr$=0.025 as thermal turbulence models, and the EBM (left) and the Launder-Sharma $k$-$\epsilon$ (right) as momentum turbulence models.}
	\label{comparison_2}
\end{figure*}

Once trained, the data-driven AHFM was verified with some non-isothermal simulations run in OpenFoam, for which the model was compared with other theory-driven closures \cite{kays1994turbulent,shams2019towards,manservisi2014cfd}. The verification highlighted the sensitivity of the data-driven AHFM to the momentum turbulence model applied in combination with it. As an example, Figure \ref{comparison_2}
presents the validation of the data-driven AHFM (denoted as ANN in the Figure) for turbulent channel flow, and the comparison with other theory-driven models such as the Manservisi model (MM) \cite{manservisi2014cfd}, the Kays correlation \cite{kays1994turbulent} and the AHFM developed by Shams (AHFM) \cite{shams2019towards}. In the figures on the left, the AHFM is coupled with the Elliptic Blending Reynolds Stress Model (EBRSM) \cite{manceau2002elliptic} while in the figure on the right the AHFM is coupled with the Launder-Sharma $k$-$\epsilon$. Note that when the Reynolds stresses are accurately modelled with second order closures, the data-driven AHFM outperforms the traditional AHFM considered for comparison \cite{kays1994turbulent,manservisi2014cfd,shams2019towards}. However, the accuracy deteriorates when the network receives Reynolds stresses modelled with the Linear Eddy Visocisty Model (LEVM), because of its dependency on the Reynolds stress anisotropy $\mathbf{b}$, defined by eq. \eqref{b_def}.

Hence, the comparison shown in Figure \ref{comparison_2} highlights that the data-driven AHFM cannot handle deficiencies of the momentum modeling. This is a significant limitation of the machine learning closure, which restricts its applicability in CFD solvers.

\begin{table}[H]
\centering
\begin{tabular}{@{}cc@{}}
\hline \hline 
\textbf{Invariant Basis} & \textbf{Tensor Basis} \\ \hline \hline \\[-1pt]
$\pi_1 = \frac{k^2}{\epsilon^2} \{ \mathbf{S}^2 \} $, $\pi_2 = \frac{k^2}{\epsilon^2} \{ \boldsymbol{\Omega}^2 \}$, & $\mathbf{T_1} = \frac{k}{\epsilon^{0.5}}\mathbf{I}$,$\mathbf{T_2} = \frac{k}{\epsilon^{0.5}}\mathbf{b},$ \\[5pt]
 $\pi_3 = \frac{1} {\{\mathbf{b}^2 \}} $, $\pi_4 = \frac{k}{\epsilon} \{ \mathbf{bS}\} $,             &      $\mathbf{T_3} = \frac{k^2}{\epsilon^{1.5}}\mathbf{S} $, $\mathbf{T_4} = \frac{k^2}{\epsilon^{1.5}}\boldsymbol{\Omega}$,         \\[5pt] 
 $\pi_5 =  \{ \mathbf{b_2} \} $ , $\pi_6 = \frac{k^2}{\epsilon^2} \{ \mathbf{bS} \boldsymbol{\Omega}\} $,               &     $\mathbf{T_5} = \frac{k^2}{\epsilon^{1.5}}\mathbf{bS}$,  $\mathbf{T_6} = \frac{k^2}{\epsilon^{1.5}}\mathbf{b}\boldsymbol{\Omega}$,            \\[5pt]
 $\pi_7 =  \frac{|| \nabla \theta||}{\sqrt{k_\theta}} \frac{k^{1.5}}{\epsilon} $ , $\pi_8= \frac{k_\theta \epsilon}{k\epsilon_\theta }$             &   $\mathbf{T_7} = \frac{k^3}{\epsilon^{2.5}} \mathbf{S}\boldsymbol{\Omega}$,           \\[5pt] 
 $Re_t = \frac{k^2}{\epsilon \nu}$, $Pr = \frac{\nu}{\alpha_l}$, & $\mathbf{T_8} = \frac{k^2}{\epsilon^{2.5}}\mathbf{bS}\boldsymbol{\Omega}$, \\[5pt]
  & \\
 \hline \hline 
\end{tabular}
\caption{Formulation of the invariant and tensor basis, computed according to \cite{fiore2022physics}. The operator $\{\cdot\}$ denotes the tensor trace. The reader is referred to the nomenclature for the detailed definition of the symbols.}
\label{tensor_basis}
\end{table}

\section{Multi-fidelity database and preliminary analyses}\label{sec2}
The database employed to train the data-driven AHFM consists of high-fidelity (DNS) data for non-isothermal turbulent channel flow \cite{kawamura2000dns,tiselj2001dns} and non-isothermal backward-facing step \cite{oder2019direct} at various Reynolds and Prandtl numbers. An additional dataset of a non-isothermal planar impinging jet \cite{duponcheel2021direct} was employed for testing. The details about the reference DNS data are provided in Table \ref{t:database}. In the present work, the training database is extended with RANS counterparts of the same flows simulated in the OpenFoam environment \cite{openfoam6}. For the RANS simulations, the Launder-Sharma $k$-$\epsilon$ model was selected as the turbulence model. The computational domains were discretized to achieve $y^+$ ranging from 0.1 to 1.0. The RANS simulations were limited to the momentum field, i.e. the temperature field was not resolved. The 
reader is referred to the Appendix \ref{setup_flows_rans} for further details about the computational setup for the training data.

\begin{table*}[t!]
\begin{center}
\begin{threeparttable}
\caption{Available DNS databases for forced convection at different $Re$ and $Pr$ numbers} \label{t:database}
\begin{tabular}{ccccc}
\hline 
\hline
Author  & Flow  & Reynolds & Prandtl   & Usage \\
  and Ref. & Configuration & number\tnote{*}  & number &  \\
\hline
\hline
 Kawamura et al.\cite{kawamura2000dns} & Channel flow & $Re_{\tau}$=180-640 & 0.025-0.71 & training/test \\
 \hline
 Tiselj et al. \cite{tiselj2001dns} & Channel flow & $Re_{\tau}$=180-590 & 0.01 & training/test \\
\hline
 Oder et al. \cite{oder2019direct}  & Backward-Facing Step & $Re_{b}$=3200 & 0.1,0.005 & training/test \\
			\hline
			Duponcheel et al. \cite{duponcheel2014assessment}  & Impining jet Flow & $Re_{\tau}=$550 & 0.031   & test \\

\hline
\hline
\end{tabular}
\begin{tablenotes}\footnotesize
\item[*] The complete definition of the Reynolds number for each flow configuration can be found in the related references.
\end{tablenotes}
\end{threeparttable}
\end{center}
\end{table*}

\begin{figure*}[t!] 
	\centering
	\includegraphics[scale=0.43]{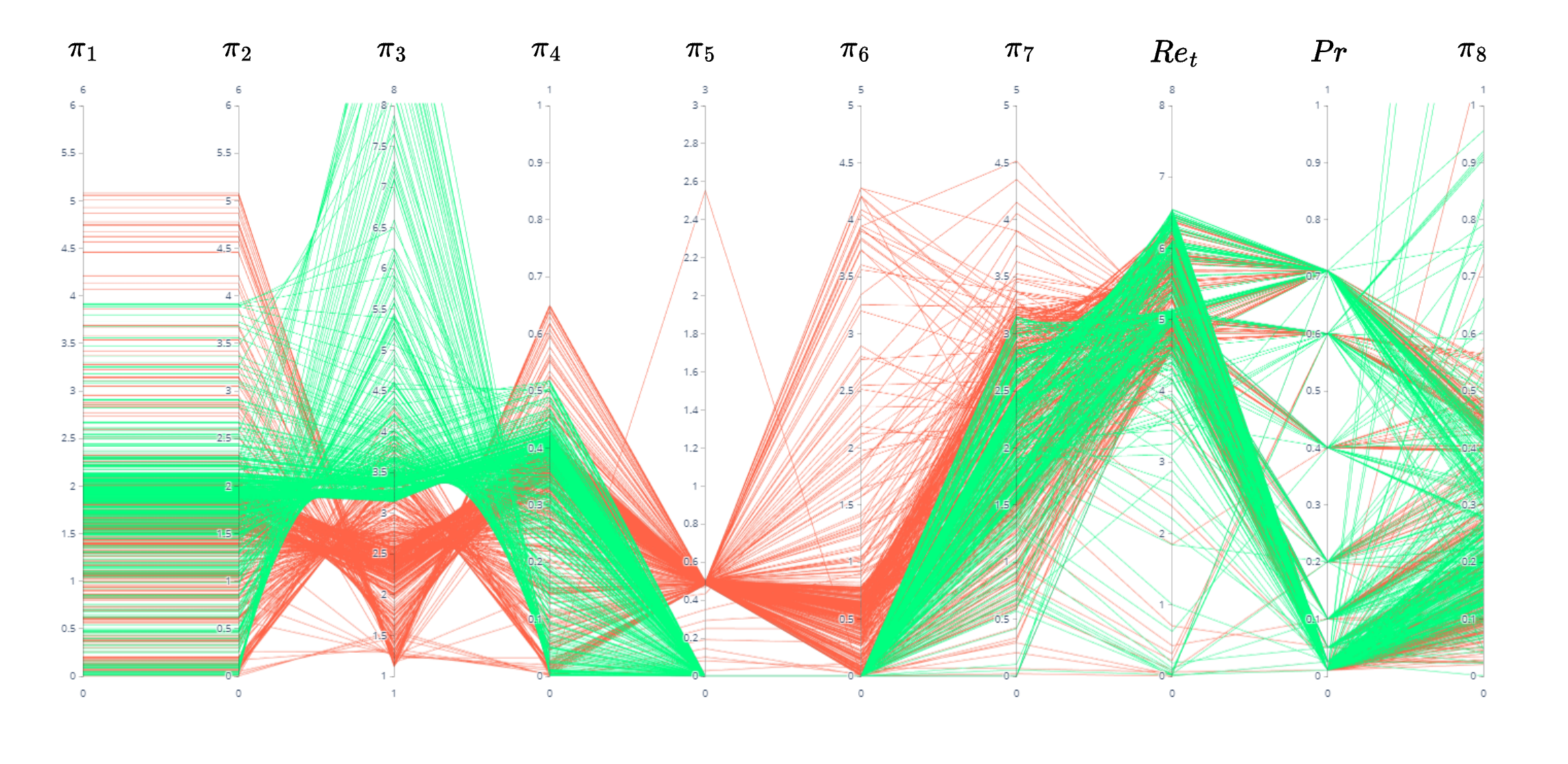}
	\caption{Visualization of the DNS (red) and RANS (green) inputs in their 10-dimensional space for non-isothermal turbulent channel flow ($Re_{\tau}=640$, $Pr=0.025$).}
	\label{f:invariants}
\end{figure*}

The collected DNS and RANS data were manipulated to compute the tensors $\mathbf{T}^i$ and the invariants $\pi_i$ indicated in Table \ref{tensor_basis}. For turbulent channel flow, the discrepancy between the $\pi_i$ computed with the two datasets is shown in Figure \ref{f:invariants}, representing the invariants in their 10-dimensional space with red (DNS) and green (RANS) lines. Specifically, each line in the plot indicates the values of the input features at each point of the computational mesh. Only 500 out of 425400 points constituting the training database are depicted for visualization purposes.

\begin{figure}[h!]
    \centering
    \includegraphics[width=0.9\linewidth]{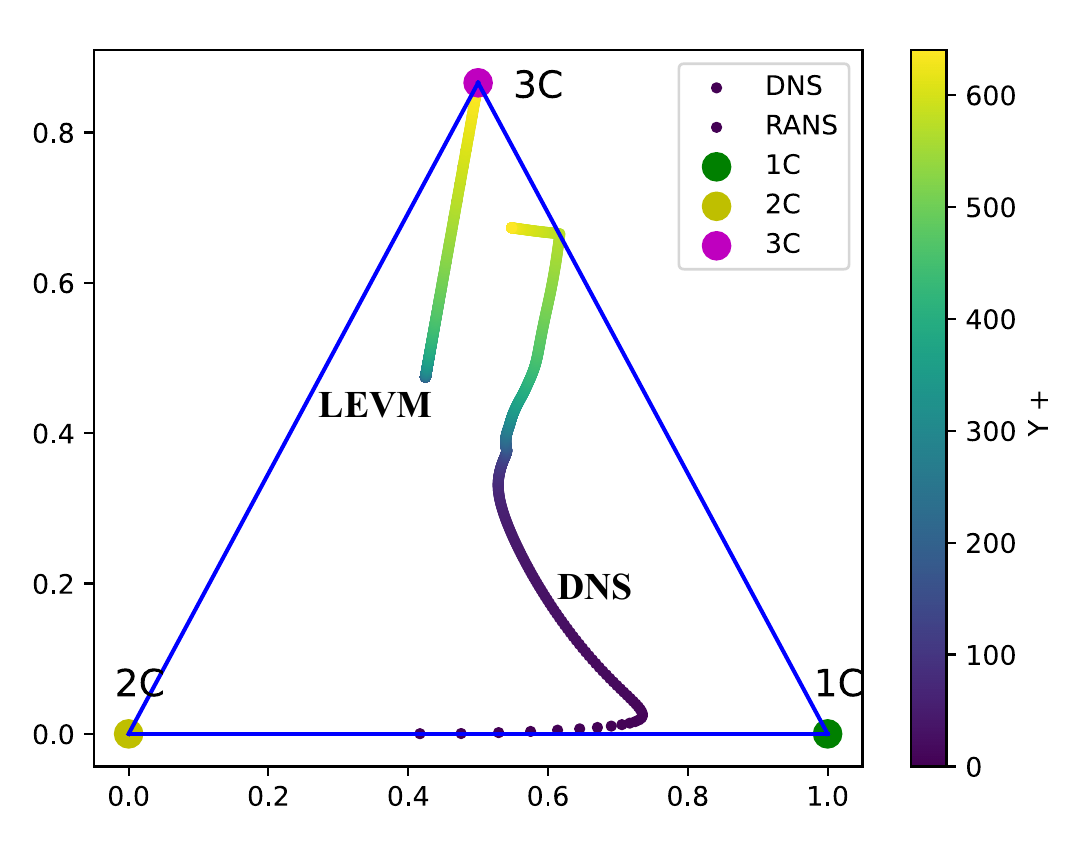}
    \caption{Anisotropic turbulence visualisation in a barycentric map in case of a turbulent channel flow. The distance between the DNS state and the Linear Eddy Viscosity Model (LEVM) can be highlighted.}
    \label{fig:BaricentricMapDist}
\end{figure}

As expected, the distribution of the invariants $\pi_1$ and $\pi_2$, $\pi_7$ and $\pi_8$ and $Re_t$ is similar between RANS and DNS since these only involve isotropic turbulent quantities and the derivatives of the mean fields ($\mathbf{U}$ and $T$). On the other hand, significant differences appear on the invariants related to the anisotropic part of the Reynolds stress tensor $\mathbf{b}$, i.e. $\pi_3$, $\pi_4$, $\pi_5$ and $\pi_6$. In fact, this behavior of the invariants reflects the differences in the tensor $\mathbf{b}$ that can be appreciated in Figure \ref{fig:BaricentricMapDist}, depicting the Reynolds stress anisotropy in the barycentric map \cite{banerjee2007presentation} for the two datasets in case of turbulent channel flow.

This representation shows that the maximum distance between the turbulent states predicted by the two approaches occurs at $y^+<30$, at which the anisotropy is severely misrepresented. The gap decreases at higher $y^+$, for which the points move towards the state of pure isotropy. 

\section{Methodology}\label{sec4}

The proposed methodology for a model handling heterogeneous input data combines dimensionality reduction, multiobjective optimization, sensitivity analysis, and uncertainty quantification. We describe the various steps in the following. Concerning the dimensionality reduction, Principal Component Analysis (PCA) was used to identify the optimal combination of features describing the different levels of fidelity as described in Section \ref{PCA}. Section \ref{mf_training} reports on the multiobjective optimization at the foundation of the model training. The performances of the model obtained with this training are analyzed and compared with the original data-driven AHFM by inspecting its output layer and estimating its sensitivities in terms of Shapley values, introduced in section \ref{shapley_sec}. Finally, the data-driven model is implemented in a CFD solver (OpenFoam), and there tested for a non-isothermal planar impinging jet at $Pr=0.01$. The details of this validation test case and the simulation setup are provided in section \ref{CFD_sec}. 

\subsection{Principal Component Analysis (PCA)}\label{PCA}

The PCA was applied on the feature matrix $\mathbf{X} \in \mathbb{R}^{N \times F}$ collecting the value of all the $F$ features on the $N$ grid points of both the training datasets. The reader is referred to \cite{bishop2006pattern} for an introduction to the PCA.

The goal was to identify $P<F$ linear combinations of the $F$ input features that explain most of the input variance and hence restrict the dimensionality of the model input's space along the associated principal directions, here denoted as $\mathbf{v}_i$, with $i=[1,..P]$. These directions are eigenvectors of the covariance matrix $\mathbf{S}$

\begin{equation}
  \mathbf{S} = \frac{1}{N-1} (\mathbf{X}-\mathbf{x}_{\mu})^T (\mathbf{X}-\mathbf{x}_{\mu}),   
\end{equation} where $\mathbf{x}_{\mu}\in\mathbb{R}^{N}$ is the vector containing the average of each column. Denoting as $\mathbf{V}=[\mathbf{v}_1\,\dots \mathbf{v}_P]\in\mathbb{R}^{N\times P}$ the matrix collecting the leading $P$ eigenvectors and $\boldsymbol{\Lambda}\in\mathbb{R}^{P\times P}$ the diagonal matrix with the associated eigenvalues sorted in decreasing order, the reduced feature space can be written as 
$\mathbf{\tilde{X}}=\mathbf{X} \mathbf{V}\in\mathbb{R}^{N\times P}$.

The correlations between the original variable and the principal components can be evaluated from the load matrix $\mathbf{L}=\mathrm{Cov}(\mathbf{X}, \tilde{\mathbf{X}}) \in \mathbb{R}^{F \times P}$. Introducing the Singular Value Decomposition of the feature matrix $\mathbf{X}=\mathbf{U}\sqrt{\boldsymbol{\Lambda}}\mathbf{V}^T$, the load matrix can be computed as defined as 

 \begin{equation}
 	\mathbf{L}=\mathrm{Cov}(\mathbf{X}, \tilde{\mathbf{X}}) = \frac{\mathbf{X}^T \tilde{\mathbf{X}}} {N-1} = \frac{\mathbf{V} \sqrt{\boldsymbol{\Lambda}} \mathbf{U}^T \mathbf{U} \sqrt{\Lambda} }{N-1} =  \frac{ \mathbf{V}\boldsymbol{\Lambda} }{N-1}.
 	\label{loading_matrix}
 \end{equation}

\subsection{Multi-fidelity training}\label{mf_training}

We denote as $\mathbf{X}_{lf}$ the feature matrix collecting data from low-fidelity simulations and as $\mathbf{X}_{hf}$ the one collecting data from high-fidelity simulations. To ensure reasonable predictions with both sets of data, the artificial neural network must be trained on the inconsistencies of the input data between the two fidelity levels. Denoting with $\mathbf{y}_{lf}$ and $\mathbf{y}_{hf}$ the predictions of the model with the inputs $\mathbf{X}_{lf}$ and $\mathbf{X}_{hf}$, and with $\hat{\mathbf{y}}_{hf}$ the reference values provided by the high-fidelity database, the cost function driving the training in case of multi-fidelity inputs is taken as:
\begin{equation}
\mathcal{L}=\mathcal{L}(\mathbf{y}_{hf},\hat{\mathbf{y}}_{hf})+\alpha \mathcal{L}(\mathbf{y}_{lf},\hat{\mathbf{y}}_{hf}),
\label{multi_loss}
\end{equation}
in which $\mathcal{L}(\cdot)$ is defined by \eqref{e:of}. This choice of the loss translates the previous regression problem introduced in section \ref{data-AHFM} into a multi-objective optimization problem in which the errors computed with both families of inputs need to be minimized. The scalar $\alpha$ in \eqref{multi_loss} is a training hyperparameter that varies between 0.001 and 1000 to build the Pareto front of the the multi-objective problem. 

\subsection{Shapley values}\label{shapley_sec}

We analyze the model sensitivity to its feature using Shapley values \cite{lundberg2017unified}. Shapley values, originating from cooperative game theory, offer a way to distribute the total gain (or cost) among the players (or features) based on their marginal contributions. For a model $f$ with input features $\mathbf{X}=(\mathbf{x}_1,\mathbf{x}_2,…,\mathbf{x}_F) $, the Shapley value $\phi_i$ for the feature $\mathbf{x}_i$ is defined as:
\begin{equation}
\phi_i = \sum_{S \subseteq K \setminus \{i\}} \frac{|S|! (|K| - |S| - 1)!}{|K|!} \left[ f(S \cup \{i\}) - f(S) \right],
\label{shapley}
\end{equation}
in which $K$ is the set including all $F$ features,  $S$ is a subset of $N$ features not containing $X_i$, $|S|$ is the cardinality of subset $S$, and $f(S)$ denotes the model prediction using the subset of features $S$. The term $f(S \cup \{i\}) - f(S)$ represents the marginal contribution of feature $\mathbf{x}_i$ when added to the subset $S$. The Shapley value calculation involves averaging these marginal contributions over all possible subsets $S$, weighted by the combinatorial factors ${|S|! (|K| - |S| - 1)!}/{|K|!}$. Note that, compared to other approaches for sensitivity analysis (e.g. the Integrated Gradient method \cite{sundararajan2017axiomatic} applied in \cite{fiore2022physics}) this method provides more complete information about the sensitivity of the model, since the discrepancies in eq. \eqref{shapley} is evaluated over the whole input space instead of targeting specific trajectories. 
Specifically, the predictions $f(S)$ for each subset are computed by averaging the model evaluations obtained by perturbing the features excluded from the subset $S$ around their predefined baseline values. In the present case, the baseline values of the ablated features is considered their average over the entire range of flows considered in the database indicated in Table \ref{t:database}. 

Note that based on eq. \eqref{shapley} the computational cost of the Shapley method is given by the product of the number of features with their number of permutations, i.e., $F \cdot F!$, which would lead to $18\cdot 18!$ model evaluations based on the list of features reported in Table \ref{tensor_basis}. To reduce the computational cost, the features were assigned to four main groups that will also ease the interpretation of the results. These groups, reported in Table \ref{tab:shapTeam}, are:  
\begin{itemize}
\item{Group 1: Momentum isotropic features (MI) depending on the velocity gradient and isotropic momentum statistics (e.g. $k$, $\epsilon$); }
\item{Group 2: Momentum anisotropic features (MA) depending on the anisotropic part of the Reynolds stress tensor $\mathbf{b}$; }
\item{Group 3: Thermal-based features (TH) depending on the molecular Prandtl number and thermal related statistics;}
\item{Group 4: The basis tensors (TE) indicated in Table \ref{tensor_basis}. }
\end{itemize}

\begin{table*}[htb!]
		\caption[Features grouping for the Shapley value analysis]{Features grouped for the Shapley value analysis. The definition of the features can be found in Table \ref{tensor_basis}. }
	\label{tab:shapTeam}
		\centering
		\begin{tabular}{@{}lccc@{}}
			\toprule
			\multicolumn{4}{c}{\textbf{Groups of features for the Shapley value analysis}}                     \\ \midrule
			\multicolumn{1}{c}{\begin{tabular}[c]{@{}c@{}} \textbf{Momentum}\\ \textbf{isotropic (MI)}\end{tabular}} & \textbf{Thermal (TH)} & \begin{tabular}[c]{@{}c@{}}\textbf{Momentum} \\ \textbf{anisotropic (MA)}\end{tabular} & \textbf{Tensors (TE)} \\ \hline \hline \\[-0.5em]
			$\pi_1$, $\pi_2$   & $Pr$, $R$           & $\pi_5$,  $\pi_3$ & $\mathbf{T}_1$, $\mathbf{T}_2$, $\mathbf{T}_3$, $\mathbf{T}_4$ \\
			$\pi_4$, $Re_\tau$ & $\pi_7$, $\nabla T$ & $\pi_6$           & $\mathbf{T}_5$, $\mathbf{T}_6$, $\mathbf{T}_7$, $\mathbf{T}_8$ \\ \bottomrule
		\end{tabular}%
\end{table*}

\subsection{Uncertainty propagation}\label{unc_sec}
The uncertainty analysis of the trained data-driven AHFMs was carried out to evaluate their robustness with respect to the momentum treatment. The method is based on the perturbation of the turbulent state in the barycentric map introduced by Figure \ref{fig:BaricentricMapDist} and, specifically, between the true state (DNS) and the one predicted by Linear Eddy Viscosity Models (LEVMs). The coordinates of the points in the map are computed from the eigenvalues $\psi_i$ ($i=1,2,3$) of the anisotropic part of the Reynolds stress tensor $\mathbf{b}$. The vertices of the triangle ($\mathbf{z}_{1C}$,$\mathbf{z}_{2C}$,$\mathbf{z}_{3C}$) represent the limiting states of turbulence:
\begin{itemize}
\item one component ($\mathbf{z}_{1C}=[1,0]$) for which $\psi_i = \frac{2}{3}, -\frac{1}{3}, -\frac{1}{3}$.
\item two components ($\mathbf{z}_{2C}=[0,0]$) for which $\psi_i = \frac{1}{6}, -\frac{1}{6}, -\frac{1}{3}$.
\item isotropic ($\mathbf{z}_{3C}=[1/2,\sqrt{3}/2]$), for which $\psi_i$ are all zero. 
\end{itemize}

Each point $\mathbf{z}\in\mathbb{R}^2$ in this plane is associated to a specific set of eigenvalues $\boldsymbol{\psi}\in\mathbb{R}^{3}$ by a linear mapping \cite{banerjee2007presentation}:

\begin{equation}
	\mathbf{z}^*= \mathbf{z}_{1C}(\psi_1-\psi_2) + \mathbf{z}_{2C}(2\psi_2-2\psi_3) + \mathbf{z}_{3C} (3\psi_3 + 1 ),
\end{equation} complemented with the conditions: 
\begin{equation}
\psi_1 + \psi_2 + \psi_3 = 0,
\end{equation}

Writing the mapping $\boldsymbol{\psi}\rightarrow\mathbf{z}$ as $\mathbf{z}=\mathbf{B}\boldsymbol{\psi}$, its inverse becomes $\boldsymbol{\psi}=\mathbf{B}^{-1}\mathbf{z}$.

\begin{figure}[H]
	\centering
	\includegraphics[width = 1.02\linewidth]{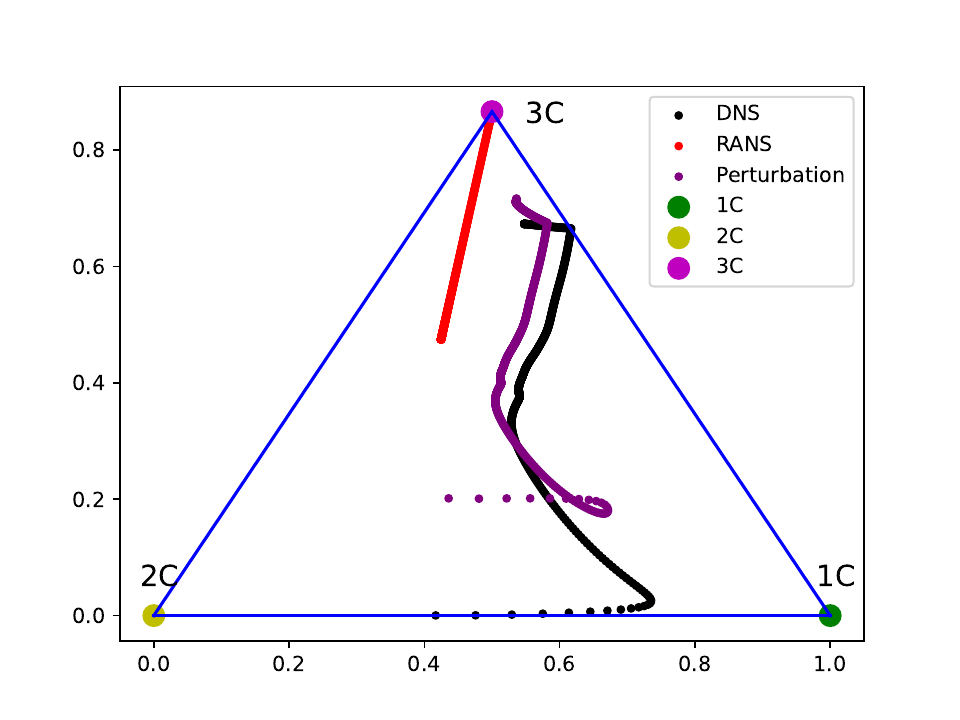}
	\caption{Perturbed anisotropy state of turbulence in the baycentric map.}
	\label{baryc_map_pert}
\end{figure}

This mapping was here used to propagate uncertainties in the baricentric map to uncertainties in the associated anisotropic stress and then the turbulent heat flux through the model. To this end, the high-fidelity (DNS) state $\mathbf{z}_{hf}$ in the barycentric map is perturbed towards the state of turbulence given by the low fidelity (RANS) momentum treatment $\mathbf{z}_{lf}$, leading to a modified location $\mathbf{z}^*$:
\begin{equation}
	\mathbf{z}^* = \mathbf{z}_{hf} + \Delta (\mathbf{z}_{lf}-\mathbf{z}_{hf}),
\end{equation}
in which $\Delta$ is a uniform random variable with range $[0,1]$. An example of a perturbed state for turbulent channel flow ($Re_{\tau}=640$) is given in Figure \ref{baryc_map_pert}.

Based on the perturbed location $\mathbf{z}^*$ in the barycentric map, the new vector of eigenvalues $\boldsymbol{\psi}^*$ is computed as
\begin{equation}
	\boldsymbol{\psi}^* = \mathbf{B}^{-1} \mathbf{z}^*\,.
\end{equation} 

The associated perturbed anisotropy tensor is then computed as:
\begin{equation}
\mathbf{b}^* = \mathbf{E} \operatorname{diag}(\boldsymbol{\psi}^*) \mathbf{E}^{-1},
\end{equation}
in which $\mathbf{E}$ is the matrix of eigenvectors of the unperturbed tensor $\mathbf{b}$. The tensors $\mathbf{T}^i$ and invariants $\pi_i$ are computed based on the perturbed $\mathbf{b}^*$ and propagated through the model.

For each point of the original DNS dataset, 50 perturbed states are sampled and submitted to the data-driven models trained in single fidelity mode \cite{fiore2022physics} and in the multi-fidelity mode described in section \ref{mf_training}. The predictions from both models are averaged, and confidence intervals are constructed to assess how uncertainties in the Reynolds stress components propagate through the data-driven closure.

\subsection{Verification test case}\label{CFD_sec}
\begin{figure}[H]
	\centering
	\includegraphics[width = 1.05\linewidth]{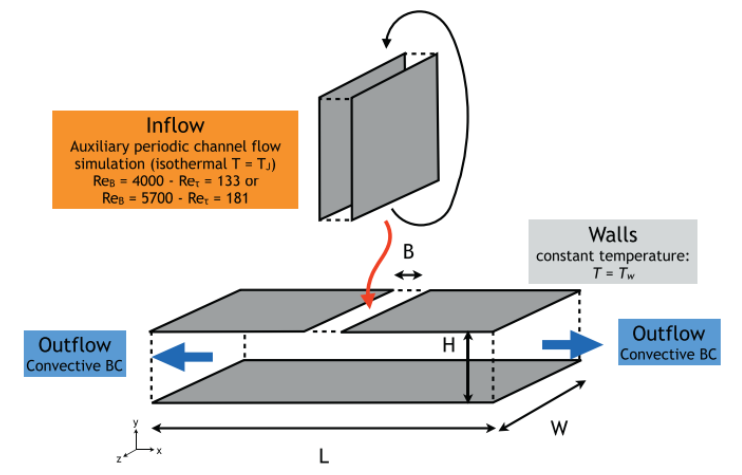}
	\caption{Schematics of the impinging jet test case developed by Duponcheel et al. \cite{duponcheel2021direct}.}
	\label{jet}
\end{figure}

To test the performance of the data-driven AHFMs generated with single fidelity \cite{fiore2022physics} and multi-fidelity training modes, the networks were implemented in a CFD solver (OpenFoam) following the procedure proposed by Maulik et al. \cite{maulik2021deploying}. The test case proposed is a non-isothermal planar impinging jet simulated by Duponcheel et al. \cite{duponcheel2021direct} at $Pr=0.01$ for the geometry depicted in Figure \ref{jet}. 

This test case is ideal for testing the data-driven closures, as this flow configuration is excluded from the training database. Additionally, for this flow configuration, the momentum field obtained with low-order turbulence models (e.g., $ k$-$epsilon$, $k$-$omega$) is affected by large uncertainties due to the extent of anisotropy, secondary recirculation, and transport effects. 

\begin{figure*}[t!]
	\centering
	\includegraphics[width = 0.8\linewidth]{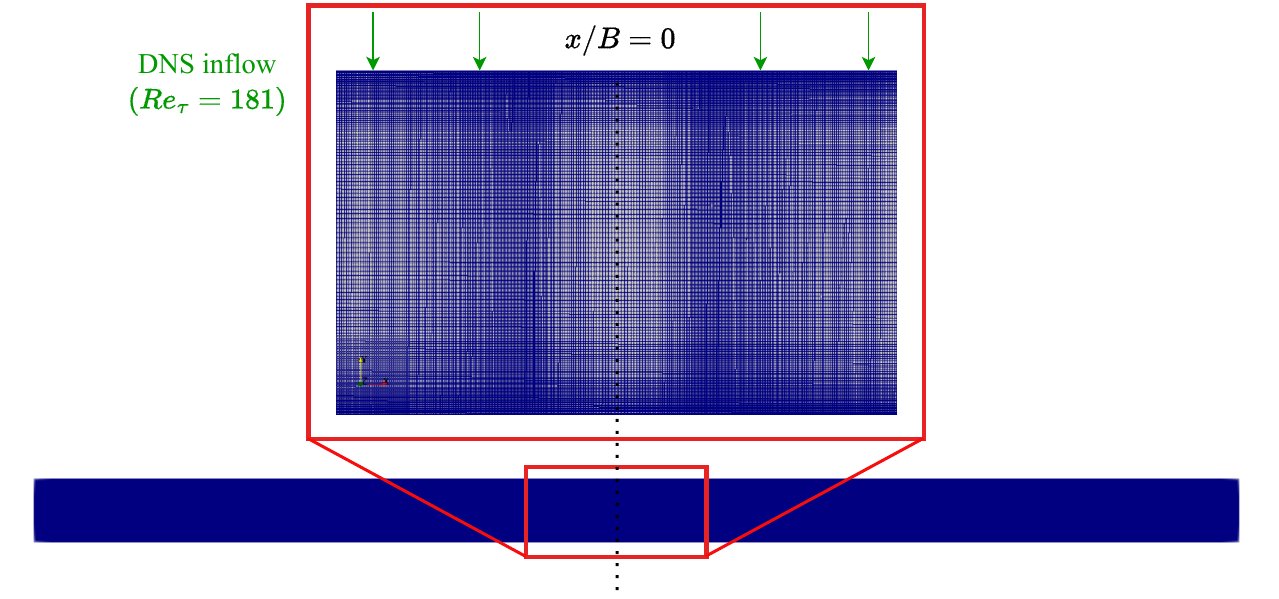}
	\caption{Computational mesh used to simulate the planar impiging jet \cite{duponcheel2021direct}.}
	\label{geom_jet}
\end{figure*}

The computational domain for this test case consists of two infinite parallel plates with a slit in the middle injecting the flow conditions reached in a fully developed turbulent channel flow at $Re_{\tau}=181$ (see Figure \ref{geom_jet}). The heat transfer is triggered by the temperature gap imposed on the walls. The structured computational grid leads to $y^+$ values ranging from 0.32 to 1.6 in the first cell near the wall, allowing a wall-resolved treatment for both momentum and thermal fields. The boundary conditions imposed are summarized in Table \ref{BC_imping}. 

\begingroup

\setlength{\tabcolsep}{6pt} 
\renewcommand{\arraystretch}{1.5}
\begin{table*}[h!]
	\caption{Overview of boundary conditions imposed for all the variables for the RANS simulation of the non-isothermal planar impinging jet. }
		\label{BC_imping}
	\centering
			\begin{tabular}{@{}llll@{}}
			\toprule
			Field      & Inlet                              & Outlet                                      & Wall                               \\ \midrule
			$\mathbf{U}$ &
			Mapped from DNS &
			$\frac{\partial U}{\partial n}=0$ or $\mathbf{U} \cdot n =0$ &
			$\mathbf{U}=0$ \\
			$T$        & Fixed: $T=2$                       & Adiabatic $\frac{\partial T}{\partial n}=0$ & Fixed: $T=0$                       \\
			$p$        & $\frac{\partial p}{\partial n }=0$ & Fixed: $p=0$                                & $\frac{\partial p}{\partial n }=0$ \\
			$k$        & Mapped from DNS                    & $\frac{\partial k}{\partial n}=0$           & $k=0$                              \\
			$\epsilon$ & Mapped from DNS                    & $\frac{\partial \epsilon}{\partial n}=0$    & $\epsilon=0$                       \\
			$k_{\theta}$ &
			$\frac{\partial k_{\theta}}{\partial n}=0$ &
			$\frac{\partial k_{\theta}}{\partial n}=0$ &
			$\frac{\partial k_{\theta}}{\partial n}=0$ \\
			$\epsilon_{\theta}$ &
			$\frac{\partial \epsilon_{\theta}}{\partial n}=0$ &
			$\frac{\partial \epsilon_{\theta}}{\partial n}=0$ &
			Manservisi wall function \\ \bottomrule
		\end{tabular}%
	
\end{table*}
\endgroup

The simulations employed the Launder-Sharma $k$-$\epsilon$ and $k-\omega$ SST as turbulent momentum closures. The energy equations was closed with the analytical model developed by Manservisi \cite{manservisi2014cfd}, the original data-driven model introduced in section \ref{data-AHFM} and the model generated with the multi-fidelity training mode.

\section{Results}\label{results}
This section presents the results of the analysis of the data-driven AHFM discussed in section \ref{data-AHFM}, comparing the model trained with multi-fidelity inputs to the one trained exclusively with DNS data. For brevity, we refer to the former as the \emph{Hybrid} ANN and the latter as the \emph{High-Fidelity} ANN.


\subsection{Results of the PCA}
The PCA was carried out on a feature matrix combining both the high-fidelity and the low-fidelity data. 

\begin{figure*} [b!]
    \centering
    \includegraphics[width = 0.6\linewidth]{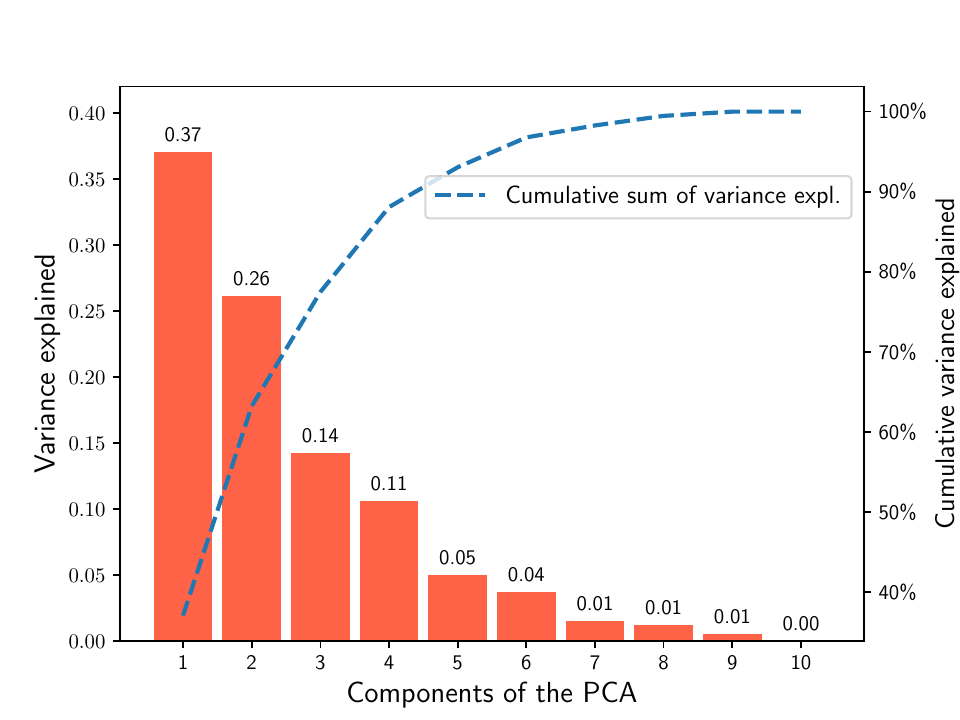}
    \caption{Variance explained for each new PC. The blue dotted line represents the cumulative sum of each PC.}
    \label{fig:PCA_CUM}
    \label{fig:PC_CumSum}
\end{figure*}

\begin{figure*}[htb!]
    \centering
    \includegraphics[width = 0.85\linewidth]{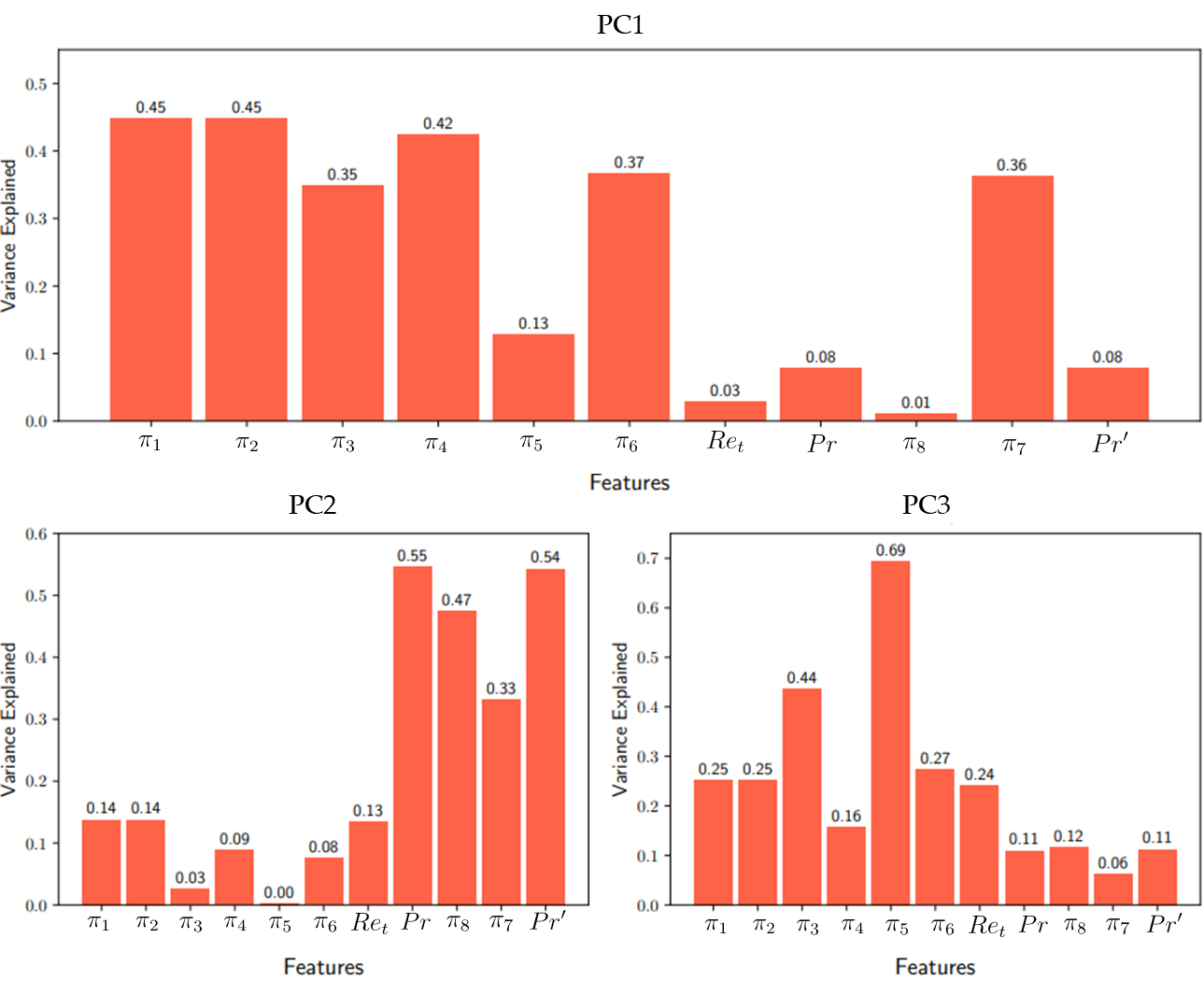}
	\caption{Features variance explained by each Principal Component.}
	\label{feature_variance}
\end{figure*}

The variance explained by each principal component is illustrated in Figure \ref{fig:PC_CumSum}, which also reports the cumulative sum of the variance explained. These results show that the first three principal components are able to explain 77.5\% of the total variance of the input data. 

The loading matrix (cf. \eqref{loading_matrix}) was computed to analyze the principal components' contribution to the original features' variance. The loadings of the first three principal components are shown in Figure \ref{feature_variance}. The loading values show that the first component (PC1) is mainly composed of momentum features depending on the mean velocity gradients and isotropic turbulence statistics (e.g., $k$, $\epsilon$), which are similar in both RANS and DNS datasets. The second principal component (PC2) is primarily constituted by thermal-based features depending on the Prandtl number of the fluid. The third principal component (PC3) depends on momentum-based features that describe the turbulence anisotropy through the deviatoric part of the Reynolds stress tensor. These latter features are misrepresented in the RANS input dataset due to the Boussinesq approximation. Interestingly, the PCA algorithm splits the entire input dataset in three clusters, including isotropic momentum-based features (PC1), thermal-based features (PC2), and anisotropic momentum-based features (PC3). 

\begin{figure*}[htb!]
    \centering
    \includegraphics[width = 0.5\linewidth]{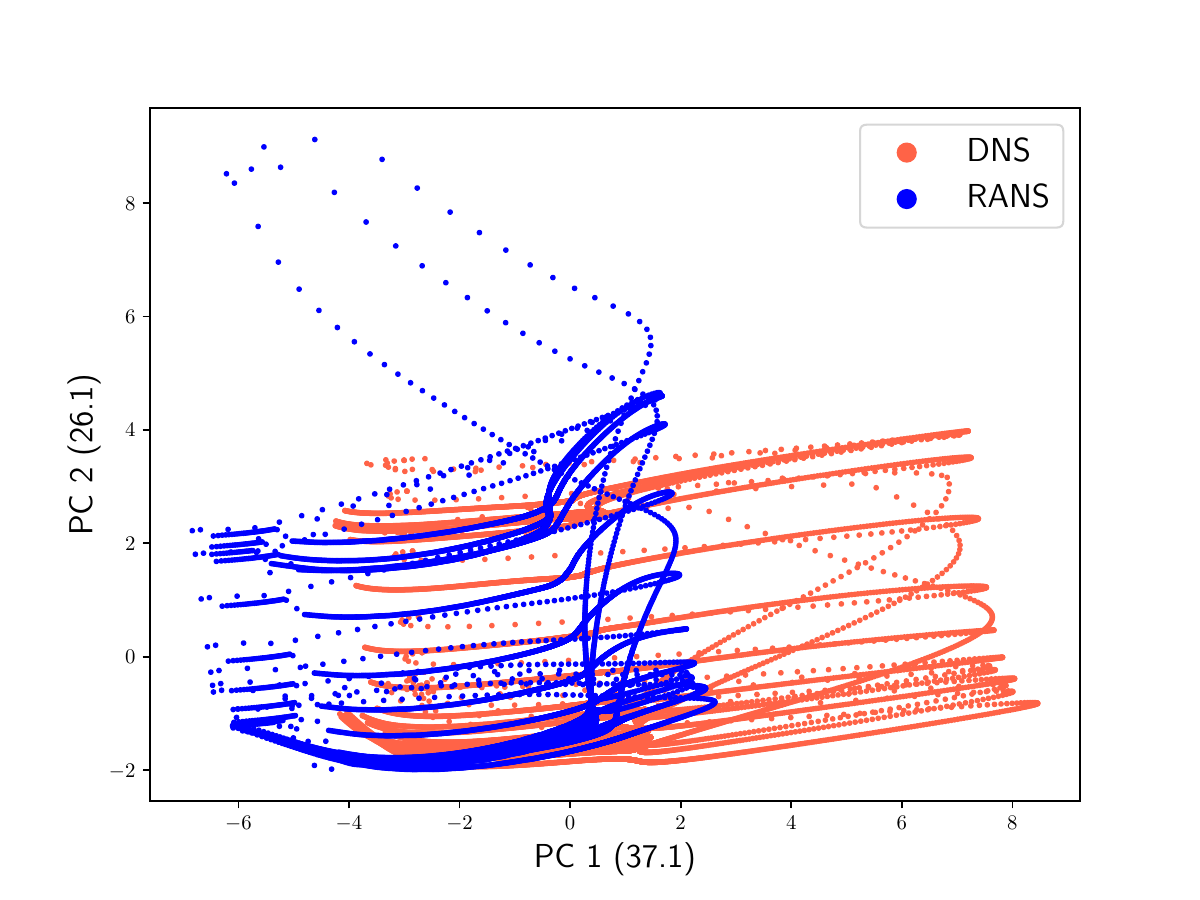}
	\caption{Projection of the RANS and DNS input data referred to non-isothermal turbulent channel flow onto the PC1 and PC2 axes.}
    \label{PC1_vs_PC2}
\end{figure*}

\begin{figure*}[htb!]
    \centering
    \includegraphics[width = 0.5\linewidth]{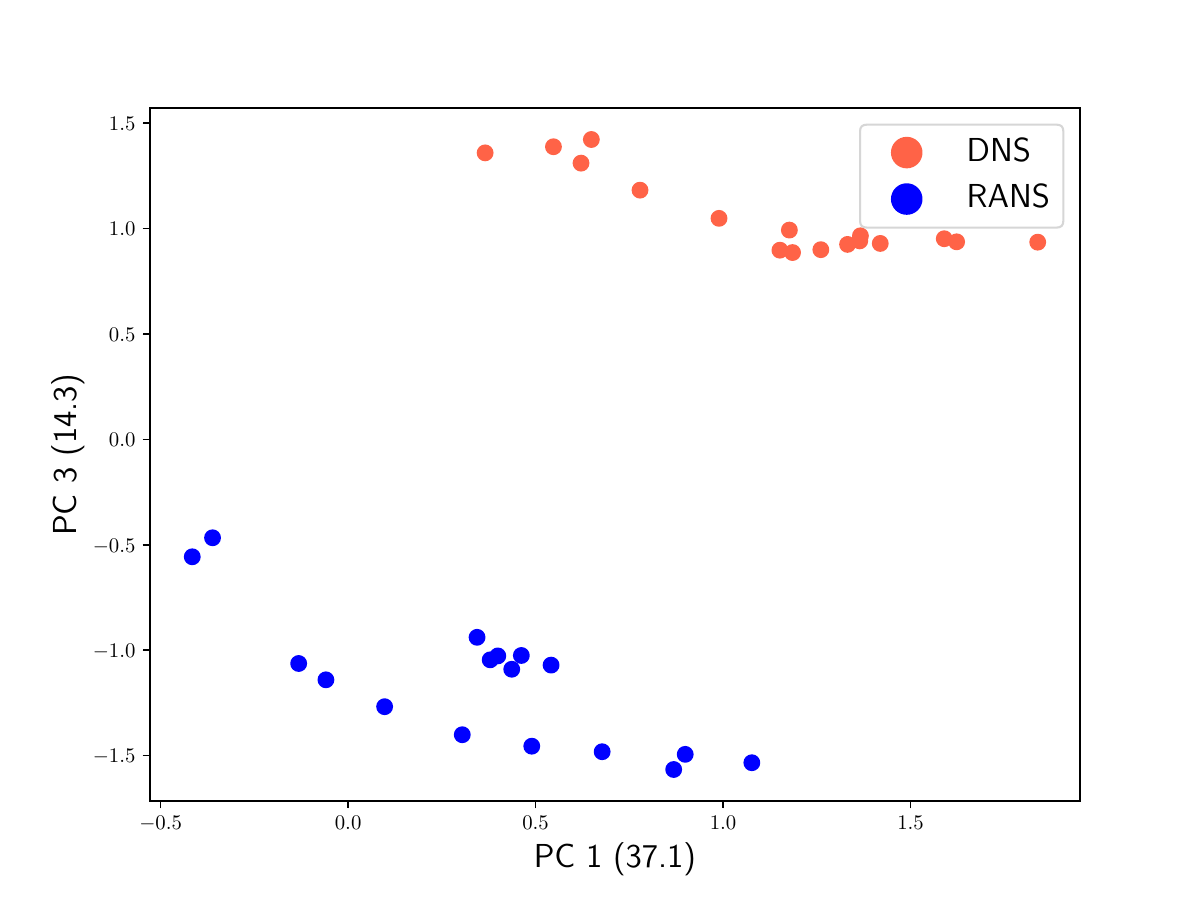}
	\caption{Projection of the RANS and DNS input data referred to non-isothermal turbulent channel flow onto the PC1 and PC3 axes.}
    \label{PC1_vs_PC3}
\end{figure*}

The two datasets (RANS and DNS) are projected onto the PCs axes and compared in Figures \ref{PC1_vs_PC2} and \ref{PC1_vs_PC3}. In particular, Figure \ref{PC1_vs_PC2} compares the two datasets in terms of PC1 and PC2. The DNS inputs achieve higher peaks of PC1 for the same PC2 values due to the higher turbulent kinetic energy retrieved in DNS simulations compared to the RANS counterpart. However, the distributions of the data on this plane are similar. A huge separation between the two datasets is observed when projected on the PC1 and PC3 plane, as done in Figure \ref{PC1_vs_PC3}. This data separation reflects the effect of the Reynolds stress modeling and suggests that the current parameter space allows the detection of the type of momentum modeling based on the distance between critical features (mainly $\pi_5$ and $\pi_3$). 
This cluster separation is crucial to the present work, as it demonstrates that specific features constituting PC3 can be utilized by a data-driven algorithm to detect the fidelity level of the momentum turbulence model and to adapt predictions accordingly.

\subsection{Performance of the training with multi-fidelity inputs}
\begin{figure*}[htb!]
	\centering
	\includegraphics[width = 0.7\linewidth]{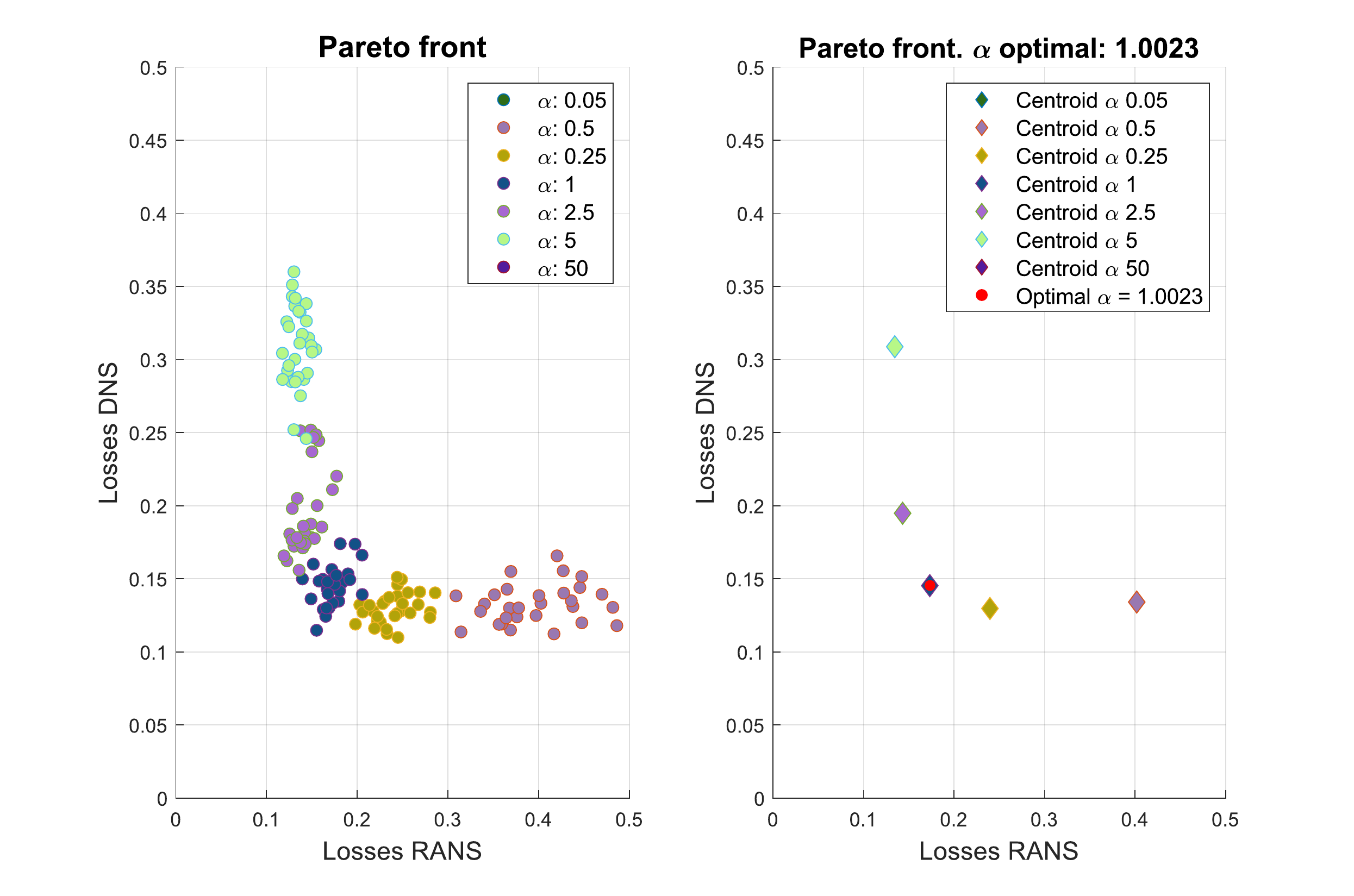}
	\caption{Pareto Front obtained by training the network 30 times for each value of $\alpha$. On the left, the entire ensemble of losses is depicted. The centroids obtained for each value of $\alpha$ are shown on the right. }
	\label{pareto_front}
\end{figure*}

The neural network architecture introduced in section \ref{data-AHFM} was trained with the multi-fidelity approach introduced in section \ref{mf_training}. Specifically, for each value of $\alpha$, 30 statistical trainings were carried out. The Pareto front of the multi-objective problem is reported in Figure \ref{pareto_front}, which is built from the final values achieved by the two losses at the end of each training. It is worth stressing that the resulting Pareto front is sharp, meaning that the training finds solutions that do not compromise the accuracy of the heat flux predictions with both types of inputs. This is an interesting result because it proves that increasing the robustness of the data-driven AHFM to the momentum modeling is possible. 

\begin{figure}[h!]
	\centering
	\includegraphics[width = 0.99\linewidth]{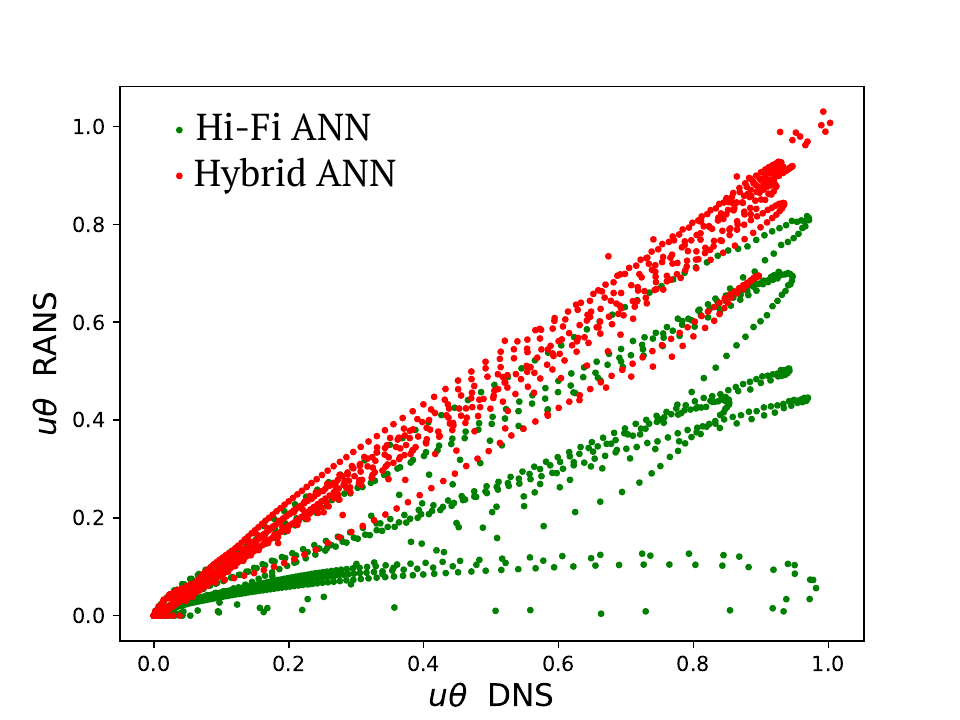}
	\includegraphics[width = 0.99\linewidth]{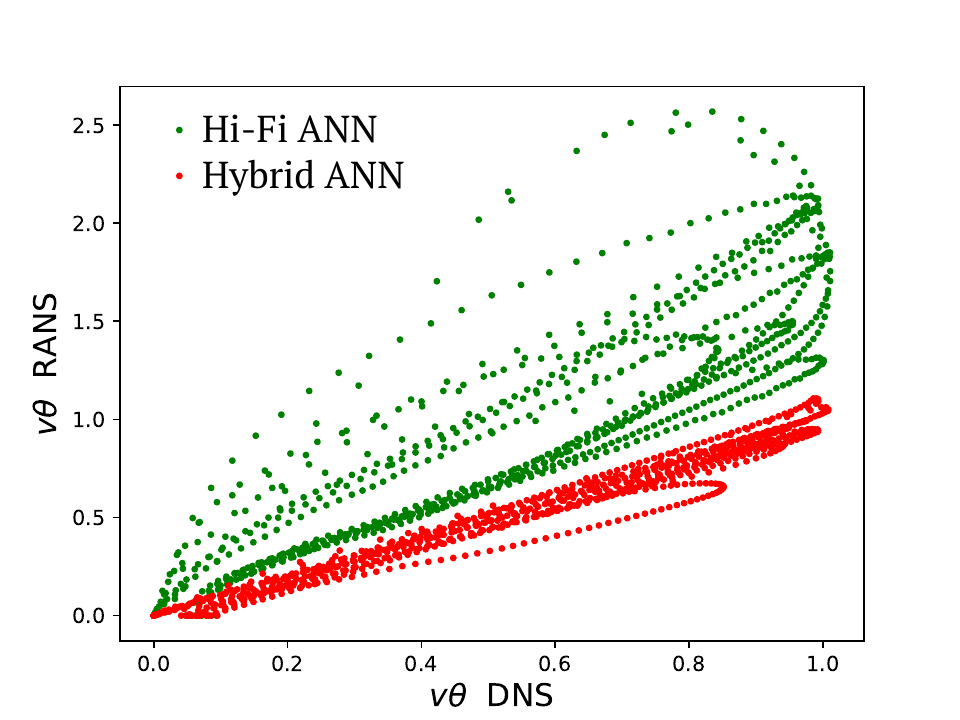}
	\caption{Comparison between the predictions with the two types of input data given by the trained ANNs, i.e. the data-driven AHFM trained with high-fidelity data ($\alpha=0$) and the same model trained with the hybrid DNS-RANS dataset ($\alpha=1.0$).}
	\label{comparison_hybrid_hifi}
\end{figure}

Regarding accuracy, a comparison between the hybrid ANN and the high-fidelity ANN is given in Figure \ref{comparison_hybrid_hifi}, which relates the network output obtained with DNS and RANS predictions. Note that hybrid ANN significantly reduces the dispersion of the heat flux predictions with the change of the momentum treatment. 

\subsection{Results of the sensitivity analysis}
\begin{figure*}[htb!]
	\begin{minipage}[H]{.49\linewidth}
		\begin{figure}[H]
			\centering
			\includegraphics[width = 0.8\linewidth ]{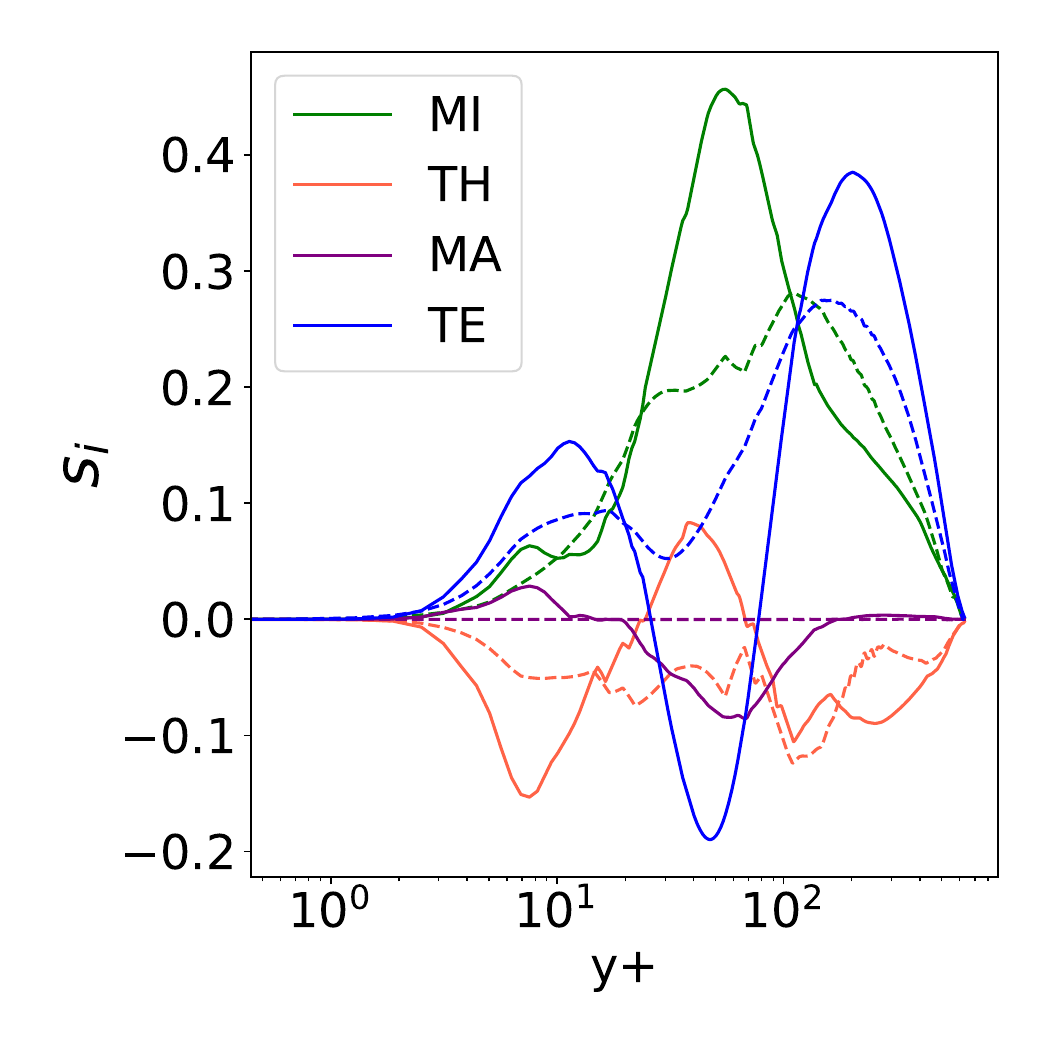}
		\end{figure}
	\end{minipage}
	\hfill
	\begin{minipage}[H]{.49\linewidth}
		\begin{figure}[H]
			\centering
			\includegraphics[width = 0.8\linewidth ]{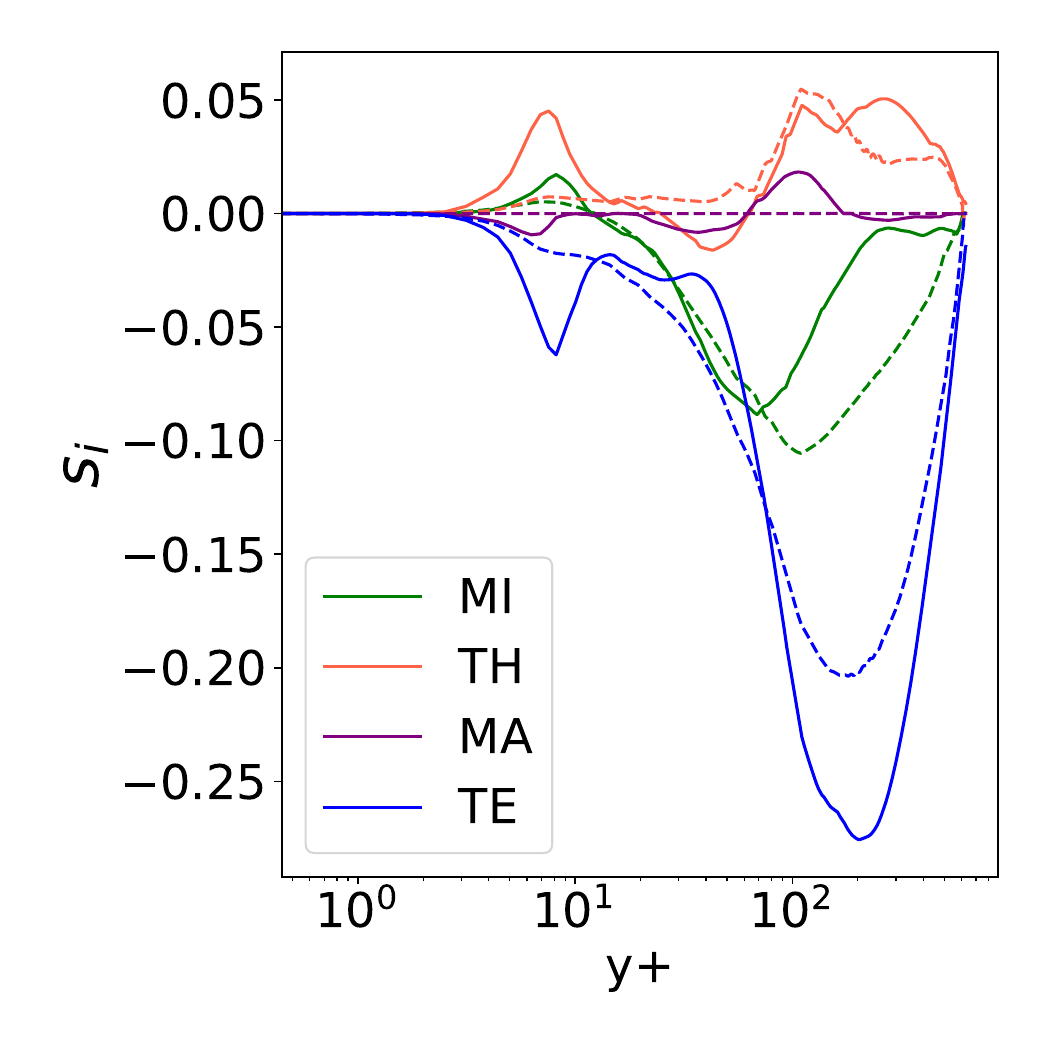}
		\end{figure}
	\end{minipage}
	\caption{Shapley values of the groups indicated in Table \ref{tab:shapTeam} for the streamwise (left) and wall normal (right) heat flux in presence of non-isothermal turbulent channel flow at $Re_{\tau}=640$ and $Pr=0.025$. The values obtained with DNS and RANS input data are indicated with solid and dashed lines, respectively. Results obtained with the hybrid network. }
	\label{shap_0025}
\end{figure*}

\begin{figure*}[htb!]
	\begin{minipage}[t]{.49\linewidth}
		\begin{figure}[H]
			\centering
			\includegraphics[width = 0.8\linewidth ]{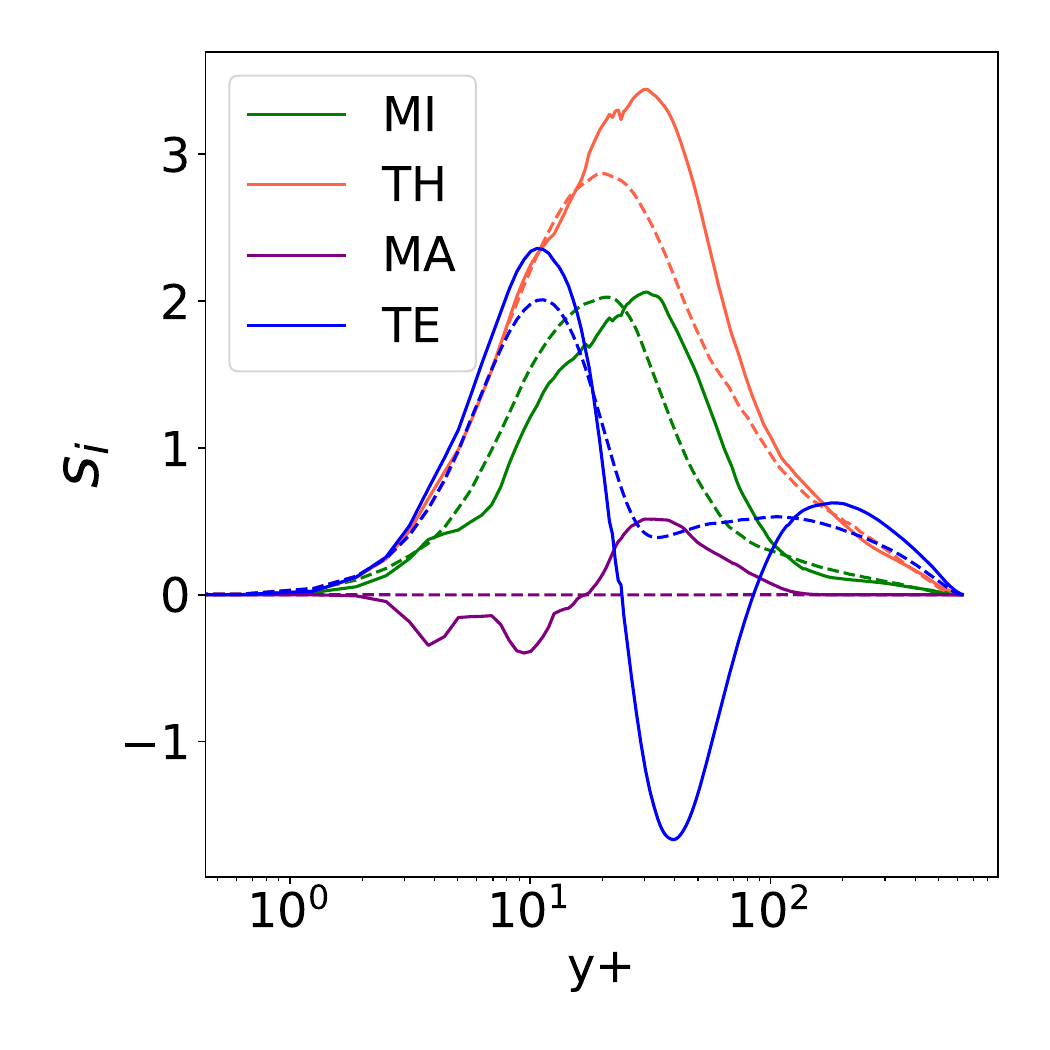}
		\end{figure}
	\end{minipage}
	\hfill
	\begin{minipage}[t]{.49\linewidth}
		\begin{figure}[H]
			\centering
			\includegraphics[width = 0.8\linewidth ]{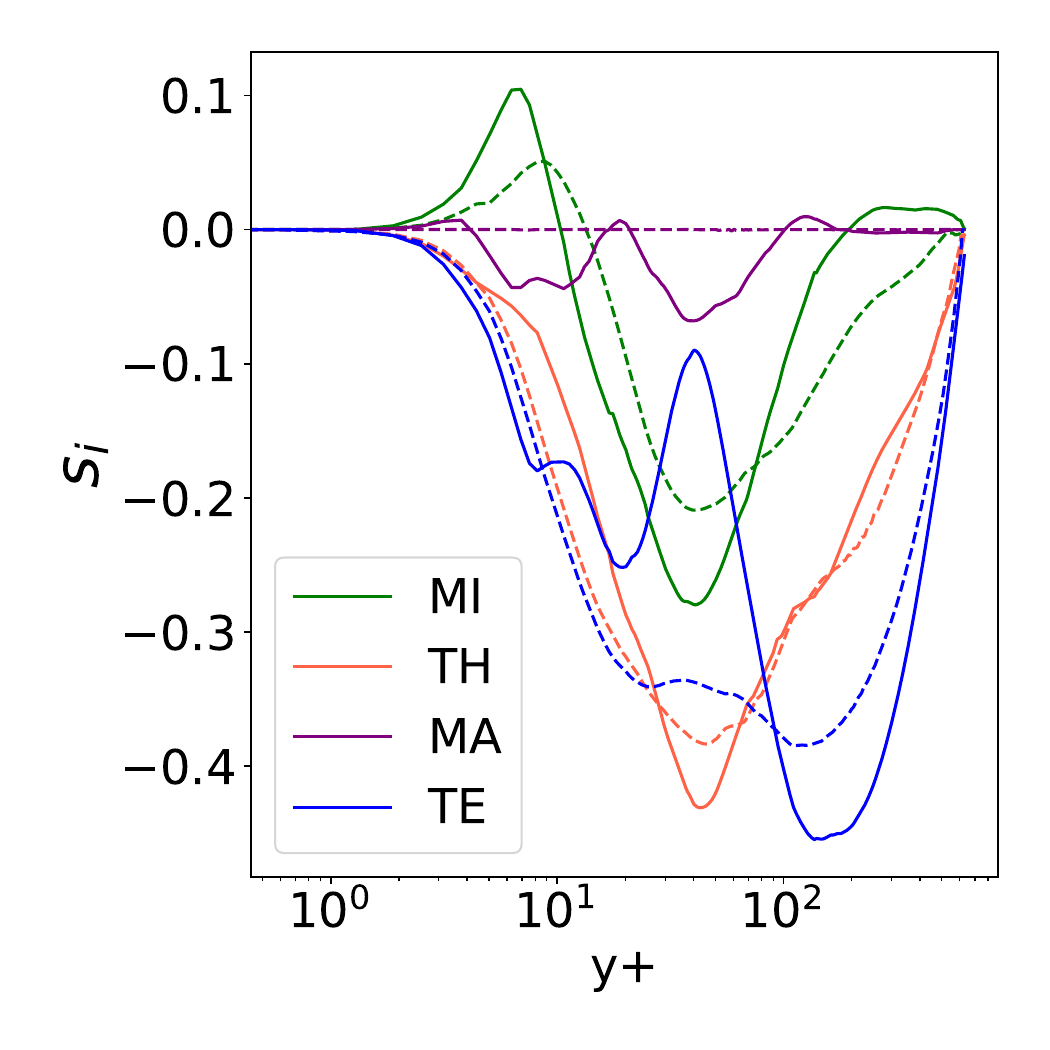}
		\end{figure}
	\end{minipage}
	\caption{Shapley values of the groups indicated in Table \ref{tab:shapTeam} for the streamwise (left) and wall normal (right) heat flux in presence of non-isothermal turbulent channel flow at $Re_{\tau}=640$ and $Pr=0.71$. The values obtained with DNS and RANS input data are indicated with solid and dashed lines, respectively. Results obtained with the hybrid network.}
	\label{shap_071}
\end{figure*}

The Shapley values defined in \eqref{shapley} were computed considering the streamwise and wall-normal components of the heat flux as the outputs of the model. 
Figures \ref{shap_071} and \ref{shap_0025} presents the obtained Shaply values for $Pr=0.71$ and $Pr=0.025$, respectively. The values obtained in the presence of DNS inputs are indicated with solid lines, of RANS inputs with dashed lines. At near unity Prandtl numbers, the data-driven model is shown to be highly sensitive to the thermal features. This sensitivity reduces at $Pr=0.025$, meaning that the dependence of the model on the thermal gradients and features increases with the Prandtl number. Clearly, the ANN trained with the multi-fidelity approach is insensitive to the anisotropy-based invariants in the presence of RANS inputs, though it preserves a small sensitivity to this group when the model is fed with high-fidelity momentum data. 

The comparison of the Shapley values in the presence of DNS and RANS data (Figure \ref{shap_071} and \ref{shap_0025}) shows that the sensitivity of the model to all the groups of features decreases when low fidelity data are detected. Hence, the hybrid ANN relies more on high-fidelity data than low-fidelity ones, i.e., the robustness in case of inaccurate Reynolds stress modeling is higher. The profiles are in fact smoother, characterized by lower peaks and fewer changes of sign. This behavior means that the model derived with multi-fidelity training simplifies when low-fidelity input data are detected. The network adapts to the uncertain inputs by reducing its sensitivity and non-linearities.

\begin{figure*}[htb!]
	\begin{minipage}[t]{.49\linewidth}
		\begin{figure}[H]
			\centering
			\includegraphics[width = 0.8\linewidth ]{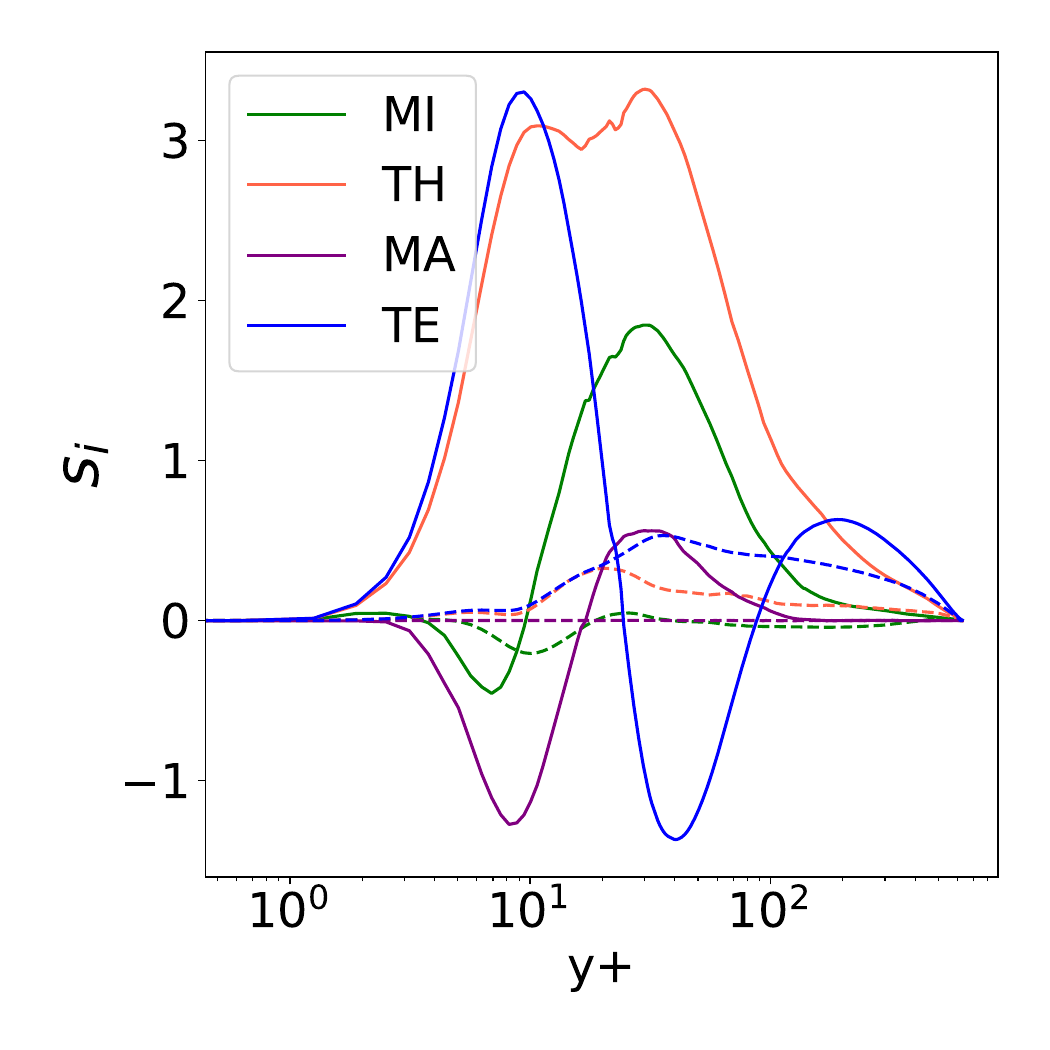}
		\end{figure}
	\end{minipage}
	\hfill
	\begin{minipage}[t]{.49\linewidth}
		\begin{figure}[H]
			\centering
			\includegraphics[width = 0.8\linewidth ]{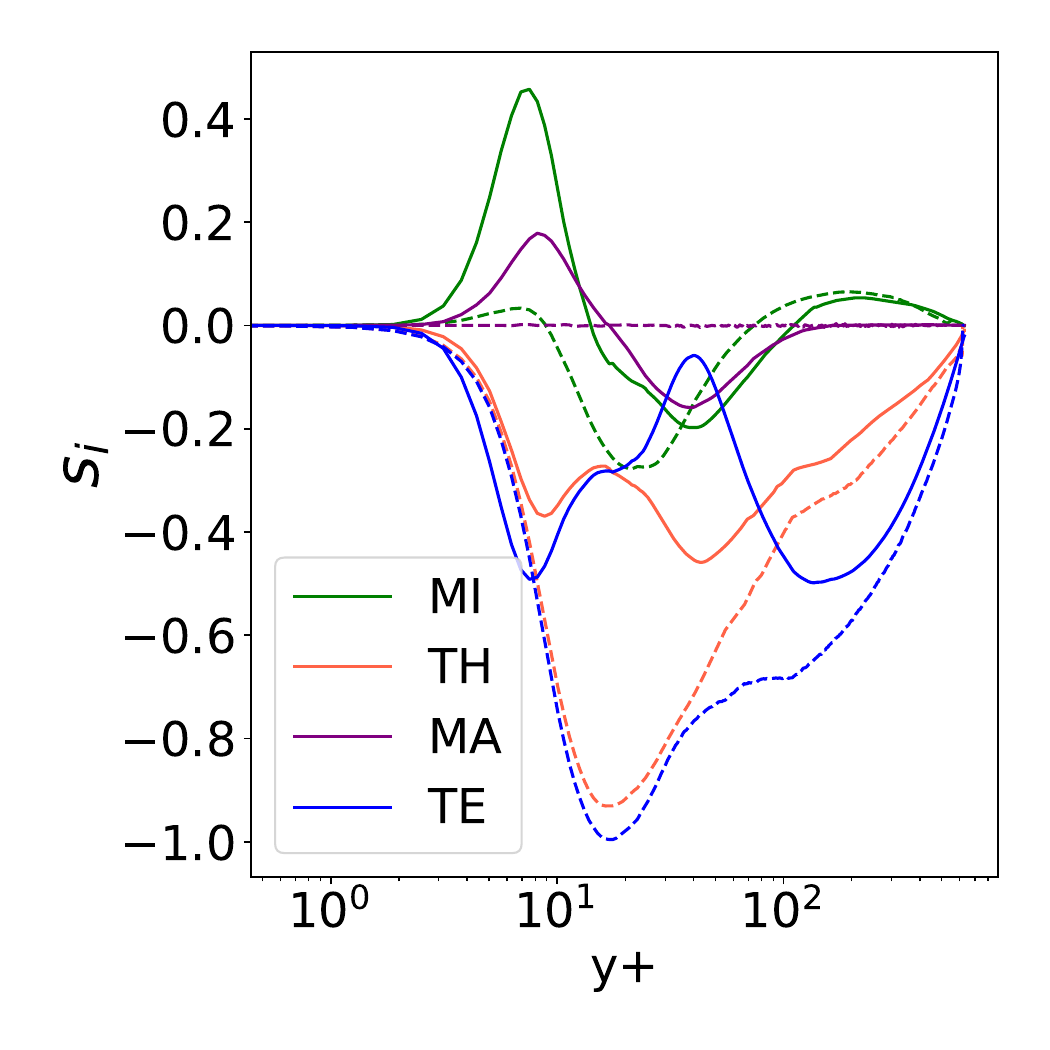}
		\end{figure}
	\end{minipage}
	\caption{Shapley values of the groups indicated in Table \ref{tab:shapTeam} for the streamwise (left) and wall normal (right) heat flux in presence of non-isothermal turbulent channel flow at $Re_{\tau}=640$ and $Pr=0.025$. The values obtained with DNS and RANS input data are indicated with solid and dashed lines, respectively. Results obtained with the Hi-Fi network. }
	\label{shap_hifi}
\end{figure*}

Figure \ref{shap_hifi} depicts the distribution of the Shapley values of the original, high-fidelity ANN for the same flow configuration at $Pr=0.71$. The comparison between Figure \ref{shap_071} and \ref{shap_hifi} highlights that the hybrid training mode reduces the sensitivity of the ANN to the features belonging to the last two groups indicated in Table \ref{tab:shapTeam}. In particular, the sensitivity drop is evident for $y^+ \approx 10$ where the momentum turbulent production reaches its peak. The barycentric triangle in Figure \ref{fig:BaricentricMapDist} shows that, at this distance from the wall, the gap between RANS and DNS turbulent states is maximum. Hence, the hybrid training mode moderates the sensitivity to the anisotropic part of the Reynolds stresses where the separation between the two categories of inputs is significant. This confirms that this training strategy effectively leads to a more robust thermal model with respect to the one trained with high-fidelity data only, which would better interface with standard momentum closures based on the eddy viscosity concept.   

\subsection{Layer output analysis}
\begin{figure*}[htb!]
	\begin{minipage}[t]{.64\linewidth}
		\begin{figure}[H][H]
	\hspace*{-0.7cm}\includegraphics[width = 0.8\linewidth]{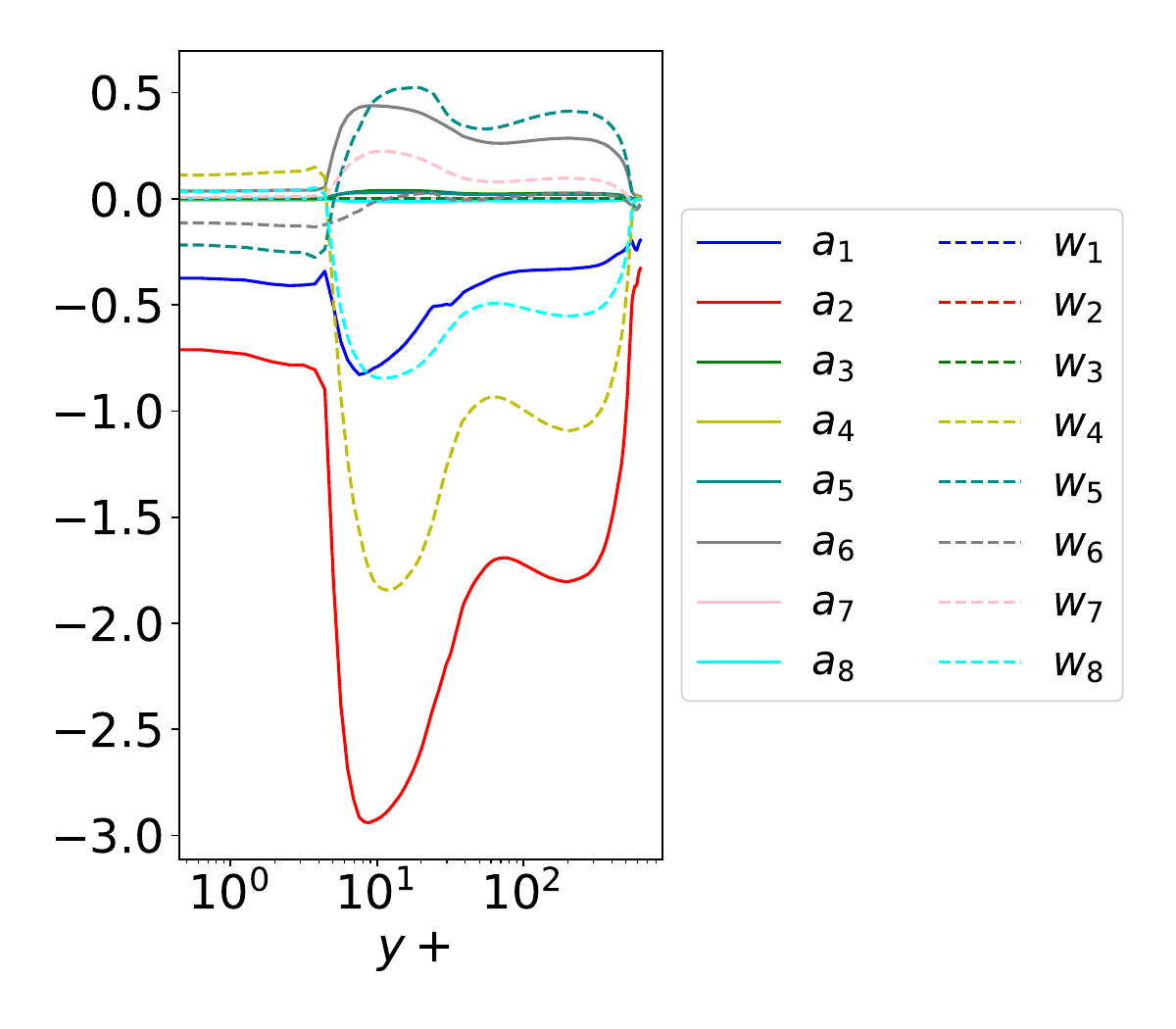}
	\hspace*{-0.7cm}\includegraphics[width = 0.8\linewidth]{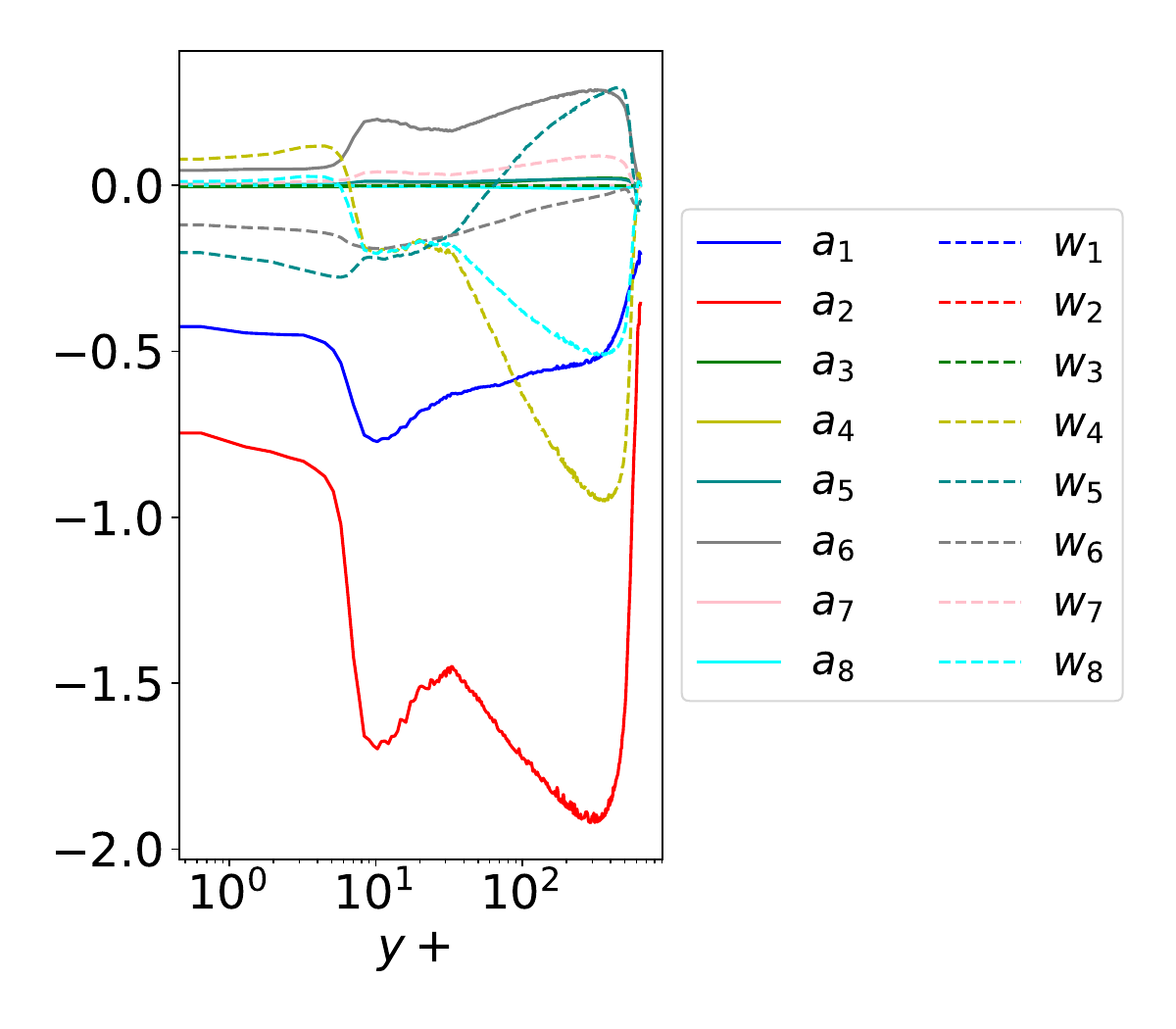}
		\end{figure}
	\end{minipage}
	\hfill
	\begin{minipage}[t]{.35\linewidth}
		\begin{figure}[H][H]
			\hspace*{-1.5cm}\includegraphics[width = 1.2\linewidth ]{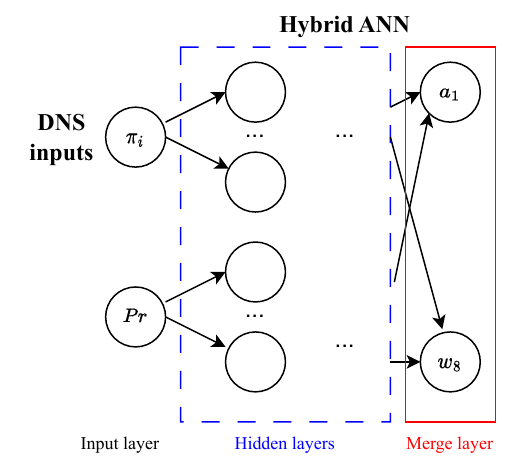} 
			
			\vspace*{0.7cm}
			\hspace*{-1.5cm}\includegraphics[width = 1.2\linewidth ]{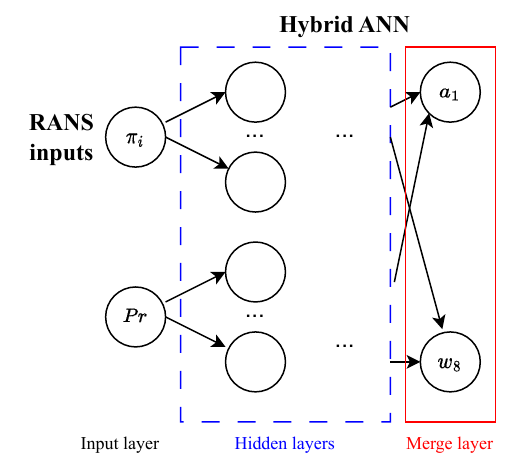}
			
			\vspace*{0.45cm}
		\end{figure}
	\end{minipage}
\caption{Output of the merge layer of the hybrid ANN for DNS (up) and RANS (bottom) inputs in case of non-isothermal turbulent channel flow at $Re_{\tau}=640$ and $Pr=0.025$.}
	\label{fig:Layer_Att_Comp}
\end{figure*}

Additional insights into the differences between the models obtained from single-fidelity and multi-fidelity training can be gained by analyzing the output layer of the network schematized in Figure \ref{f:structure}. Figure \ref{fig:Layer_Att_Comp} depicts the coefficients $a_i$ and $w_i$ at the output of the merge layer (see Figure \ref{f:structure}) of the ANN trained with the multi-fidelity approach, in case of non-isothermal turbulent channel flow at $Re_{\tau}=640$ and $Pr=0.025$ and with high-fidelity and low-fidelity momentum input data.\footnote{Note that the sign of the coefficients $a_i$ does not alter the final output, since $\mathbf{A}=\sum a_i \mathbf{T}^i$ is the Cholesky factorization of the symmetric part of $\mathbf{D}$, as indicated by eq.\eqref{chol}.}. For both kinds of input data, the dominant coefficients are $a_1$, $a_2$, $a_6$, $w_4$, and $w_6$. For $y^+>100$, the values of the predicted coefficients are similar for both DNS and RANS inputs. At $y^+\simeq 10$, where the turbulence anisotropy reaches its peak, the predicted coefficients are much higher in the presence of DNS input data than of RANS inputs. This explains the lower sensitivity to the RANS inputs detected by the Shapley value analysis in section \ref{shapley_analysis}. The reduction is significant, especially for $a_2$, the coefficient multiplying the tensor $\mathbf{T}_2$ that is proportional to the anisotropic part of the Reynolds stress tensor $\mathbf{b}$, as reported in Table \ref{tensor_basis}. 

\begin{figure*}[htb!]
	\begin{minipage}[t]{.64\linewidth}
		\begin{figure}[H][H]
			\hspace*{-0.7cm}\includegraphics[width = 0.8\linewidth]{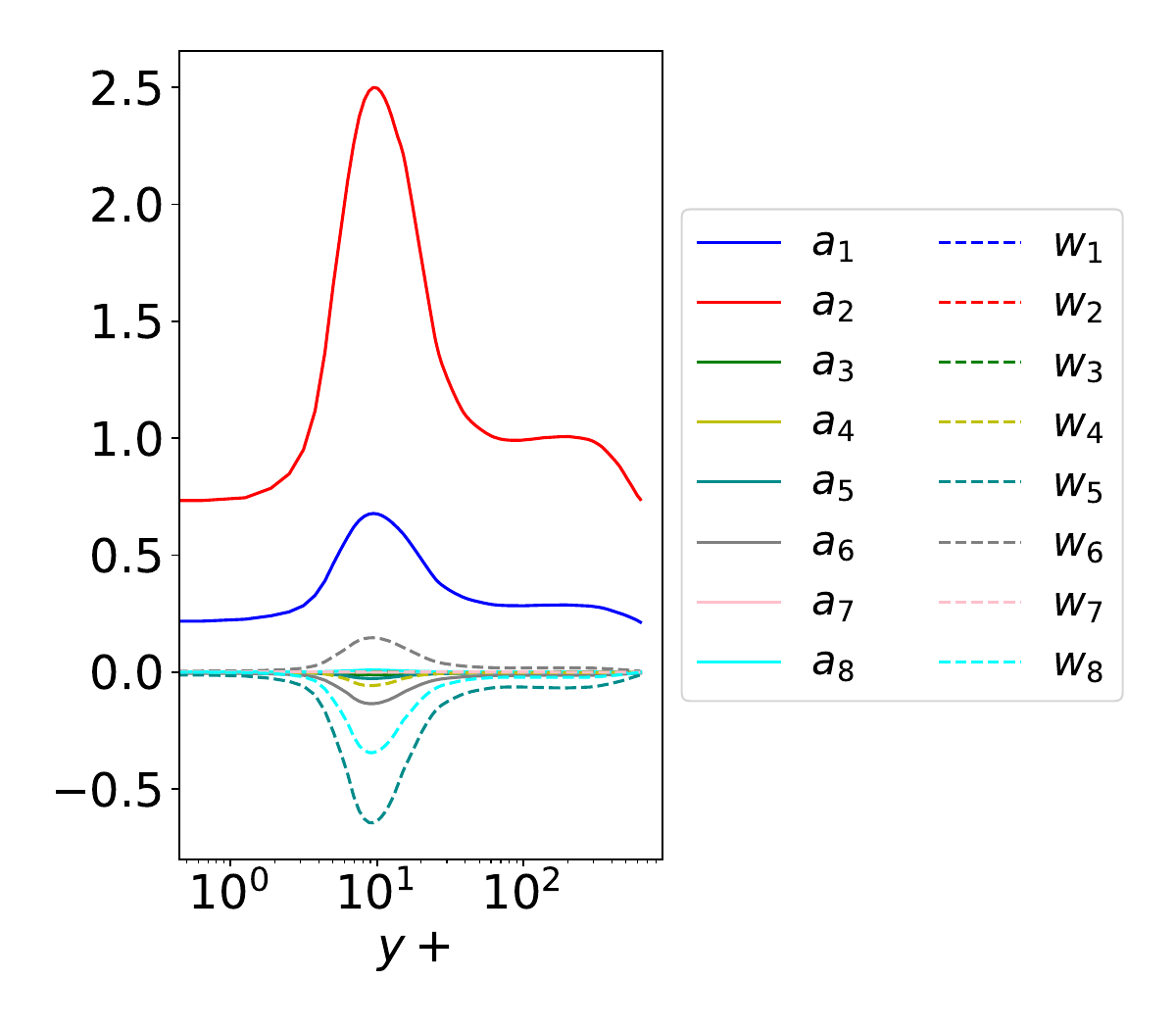}
			\hspace*{-0.7cm}\includegraphics[width = 0.8\linewidth]{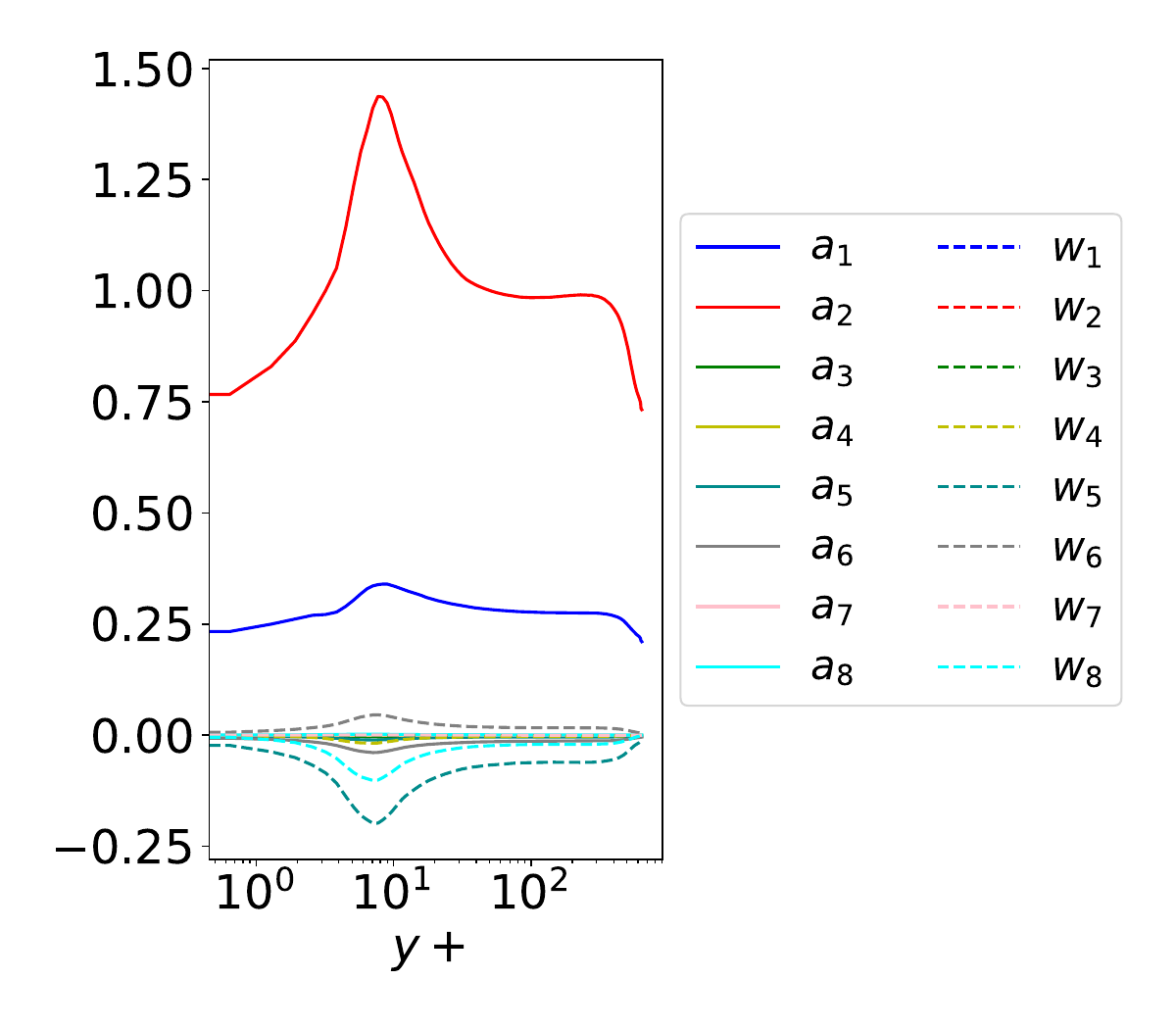}
		\end{figure}
	\end{minipage}
	\hfill
	\begin{minipage}[t]{.35\linewidth}
		\begin{figure}[H][H]
			\hspace*{-1.5cm}\includegraphics[width = 1.2\linewidth ]{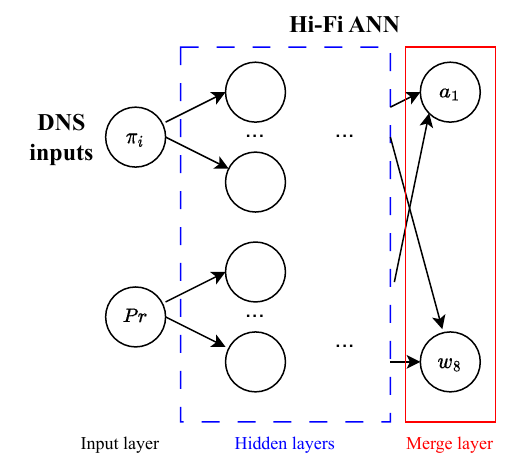} 
			
			\vspace*{0.7cm}
			\hspace*{-1.5cm}\includegraphics[width = 1.2\linewidth ]{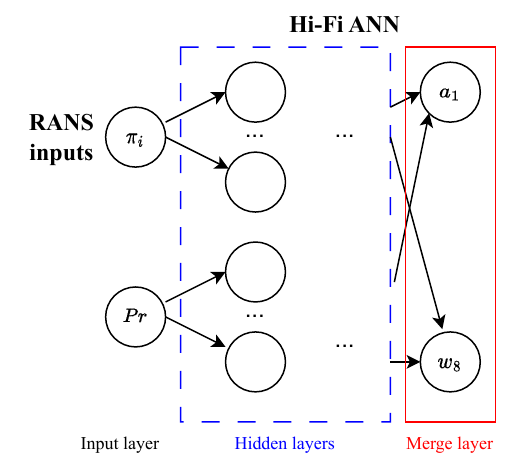}
			
			\vspace*{0.45cm}
		\end{figure}
	\end{minipage}
\caption{Output of the merge layer of the Hi-Fi ANN for DNS (up) and RANS (bottom) inputs in case of non-isothermal turbulent channel flow at $Re_{\tau}=640$ and $Pr=0.025$.}
	\label{fig:Layer_Att_Comp_alpha0}
\end{figure*}

The output of the last layer of the high-fidelity network is significantly different, as shown in Figure \ref{fig:Layer_Att_Comp_alpha0}. The coefficients $a_1$ and $a_2$ dominate above all the other terms and achieve a single peak in the region of maximum momentum turbulence production ($y^+ \approx 10$). Hence, compared to the high-fidelity network, the new model increases the dependence on $\mathbf{T}_4$,  $\mathbf{T}_6$, and  $\mathbf{T}_8$ which depend on the strain and rotation tensors. This implies that training the network with data of multiple fidelity replaces the dependence on the true Reynolds stress tensor with linear and quadratic functions of the velocity gradient. In other words, the network tries to reconstruct the true Reynolds stress anisotropy as a function of the mean velocity gradient and its uncertain estimate given by the combined momentum turbulence model. This explains the low dispersion of the predictions shown in Figure \ref{comparison_hybrid_hifi} with both kinds of momentum treatments. 

\subsection{Propagated uncertainties}
\begin{figure*}[h!]
  \centering
	\subfloat[Hybrid ANN]{%
		\label{val_val}
		
  \includegraphics[scale=0.65]{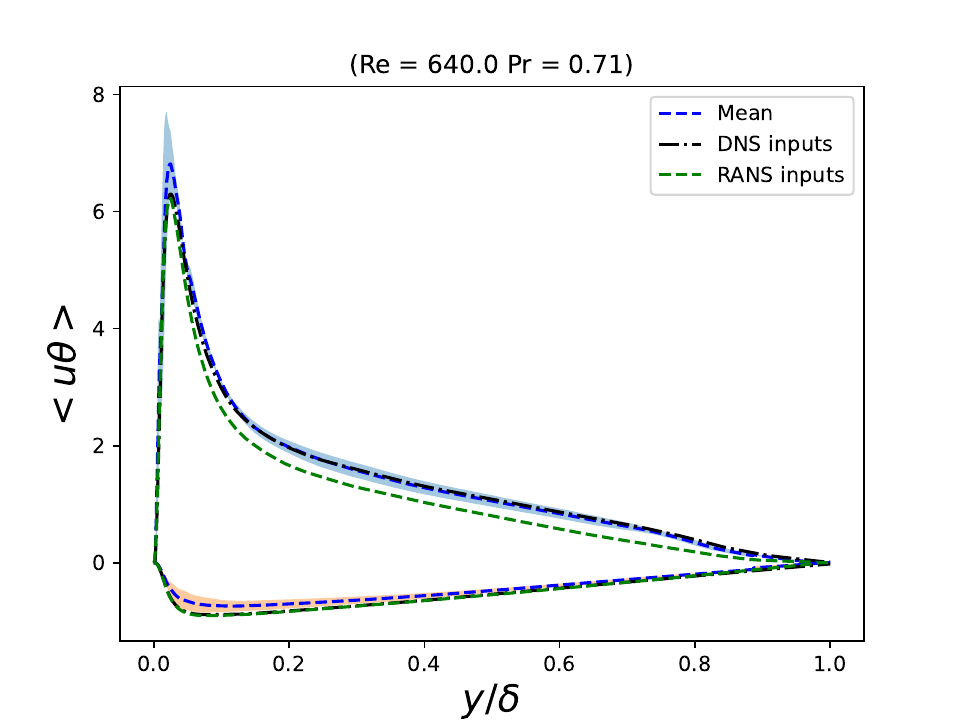}    \hspace{0.001\linewidth}}
	\subfloat[Hi-Fi ANN]{%
		\label{beta_trend}
		\includegraphics[scale=0.65]{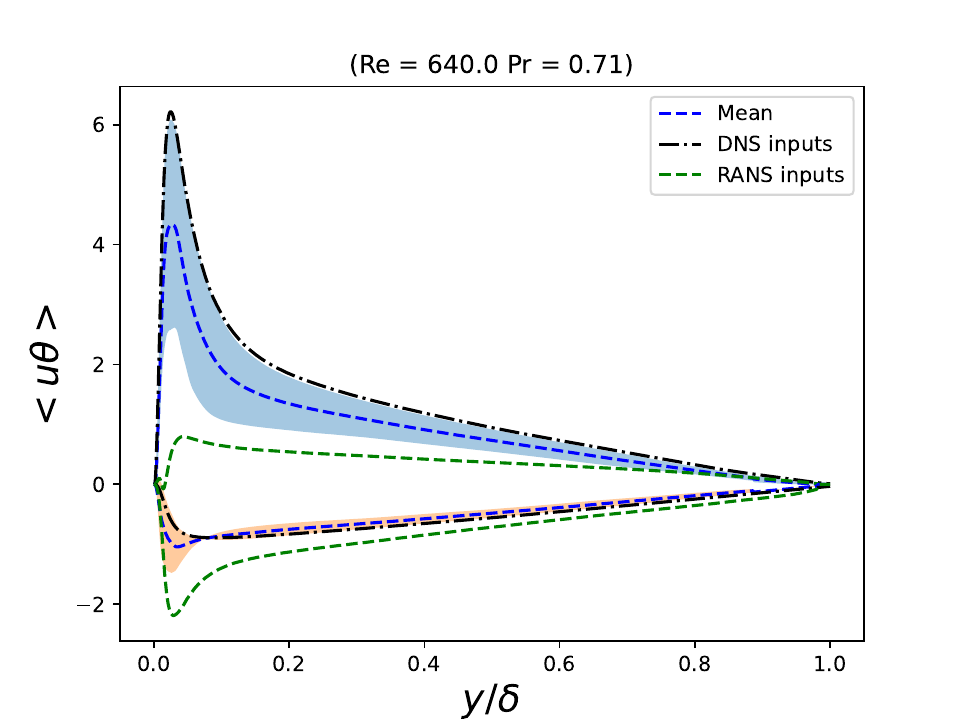}}
	\caption{Predictions of the turbulent heat flux components given by the hybrid and Hi-Fi ANNs with both the kind of inputs for non-isothermal turbulent channel flow at $Re_{\tau}=640$ and $Pr=0.71$. The model predictions' confidence intervals ($2\sigma$) due to the model-data inconsistency are indicated with shaded areas.}
	\label{unc_Pr071}
	
\end{figure*}

\begin{figure*}[htb!]
  \centering
	\subfloat[Hybrid ANN]{%
		\label{val_val}
\includegraphics[scale=0.62]{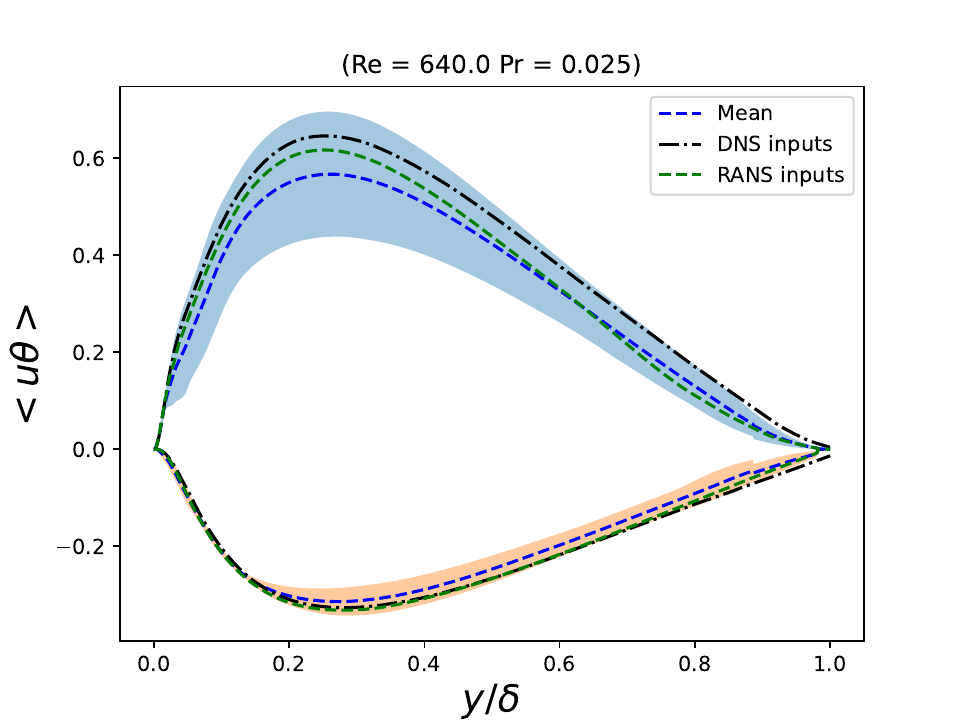}    \hspace{0.001\linewidth}}
	\subfloat[Hi-Fi ANN]{%
		\label{beta_trend}
		\includegraphics[scale=0.62]{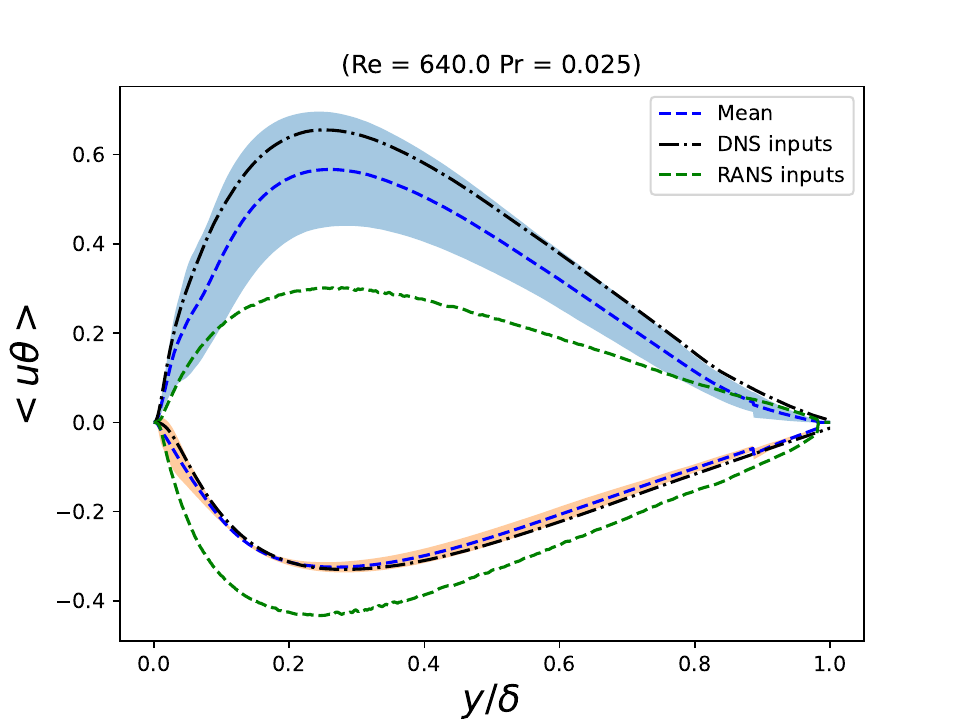}}
	\caption{Predictions of the turbulent heat flux components given by the hybrid and Hi-Fi ANNs with both the kind of inputs for non-isothermal turbulent channel flow at $Re_{\tau}=640$ and $Pr=0.025$. The model predictions' confidence intervals ($2\sigma$) due to the model-data inconsistency are indicated with shaded areas.}
	\label{unc_Pr0025}
	
\end{figure*}

The uncertainties of the Reynolds stress tensor caused by the inaccuracies of the combined momentum treatment were propagated to the output of the data-driven AHFM with the method explained in section \ref{unc_sec}. The statistics obtained for $Re_{\tau}=640$ and $Pr=0.71$ and 0.025 are shown in figures \ref{unc_Pr071} and \ref{unc_Pr0025}, respectively. The confidence intervals, highlighted with shaded areas, quantify the uncertainty of the predictions due to the inconsistency between reference and modeled input data. The comparison between the hybrid ANN and the high-fidelity ANN reveals a significant reduction of the model uncertainty when the training is enriched with RANS data. This is true at both near unity and low Prandtl numbers, albeit more evident for $Pr=0.71$, at which a stronger dependence of the heat flux field on the Reynolds stresses is expected.

\begin{figure*}[htb!]
  \centering
	\subfloat[Streamwise, $Pr=0.71$]{%
		\label{val_val}
		\includegraphics[scale=0.24]{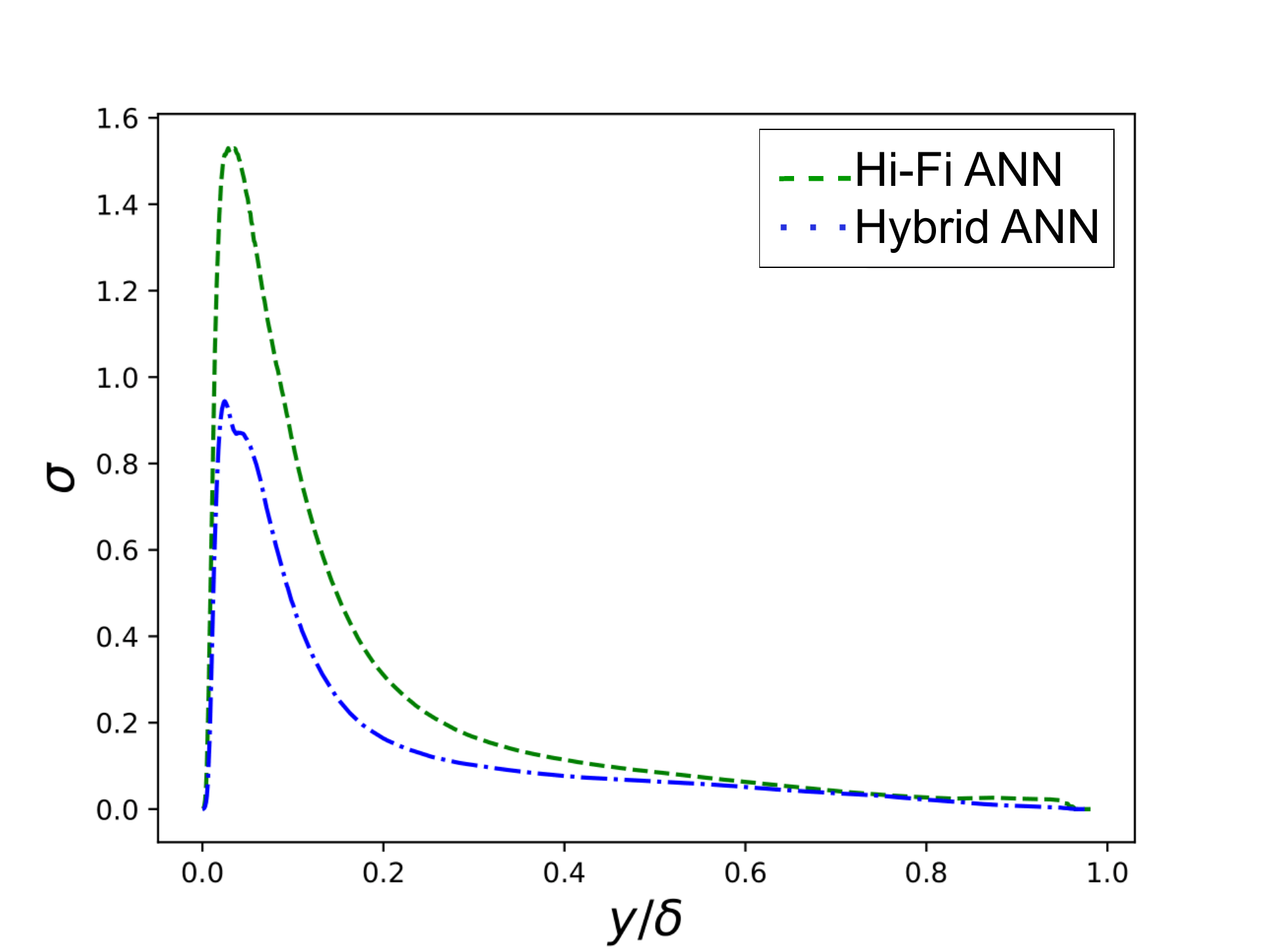}    
		\hspace{0.001\linewidth}}
	\subfloat[Wall normal, $Pr=0.71$]{%
		\label{beta_trend}
		\includegraphics[scale=0.24]{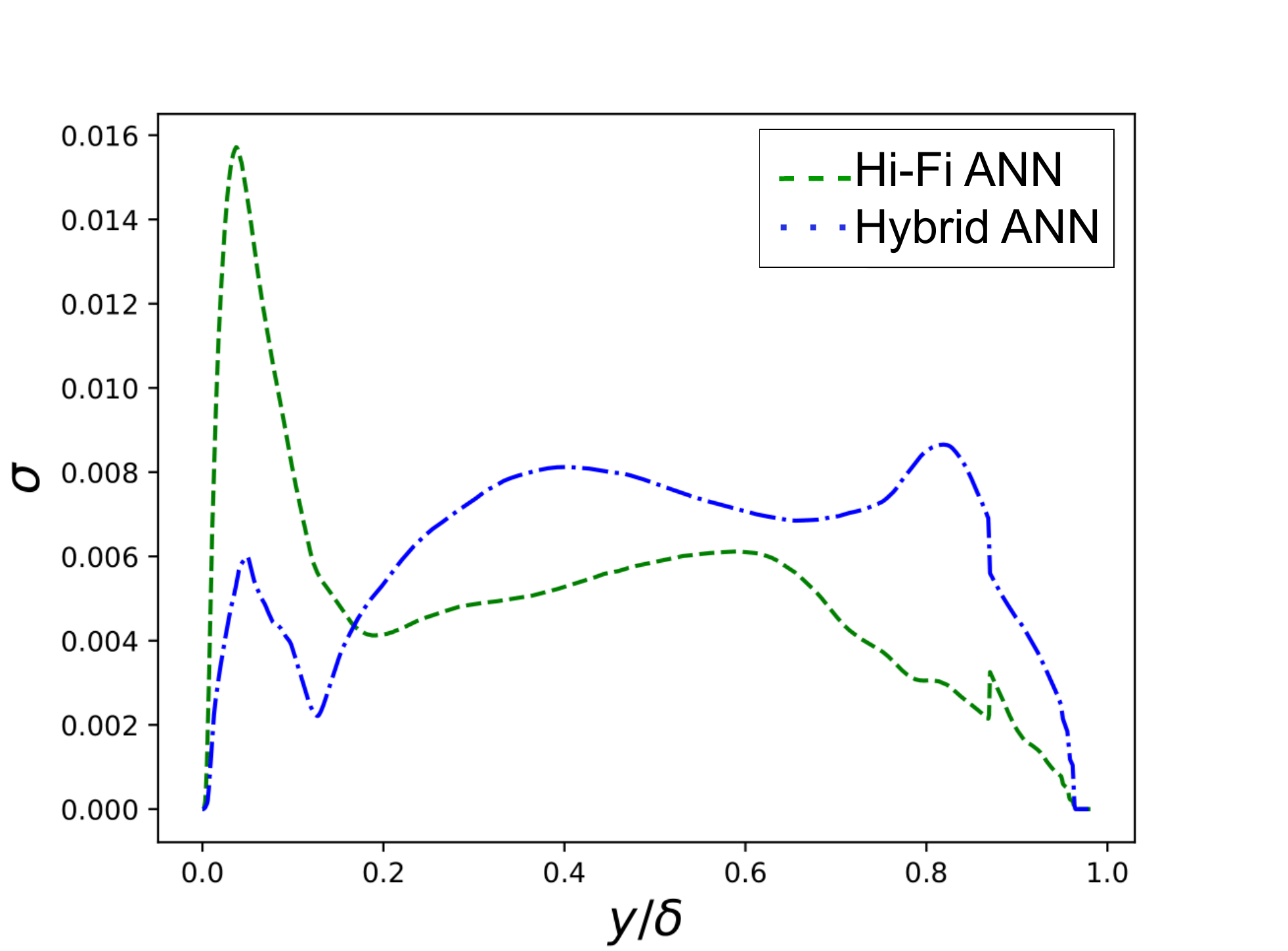}}\\
	\subfloat[Streamwise, $Pr=0.025$]{%
		\label{val_val}
		\includegraphics[scale=0.24]{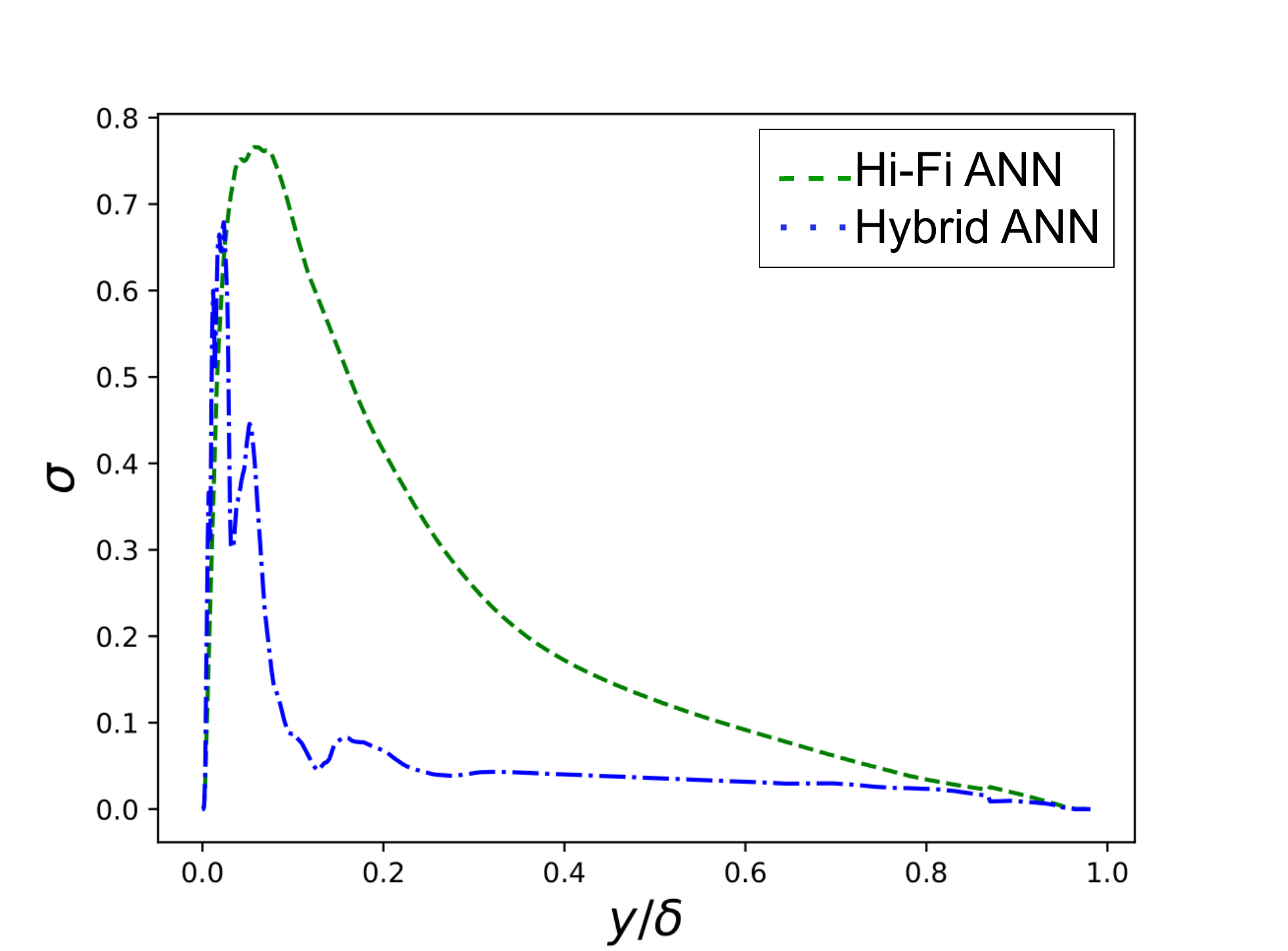} 
		   \hspace{0.001\linewidth}}
	\subfloat[Wall normal, $Pr=0.025$]{%
		\label{beta_trend}
		\includegraphics[scale=0.24]{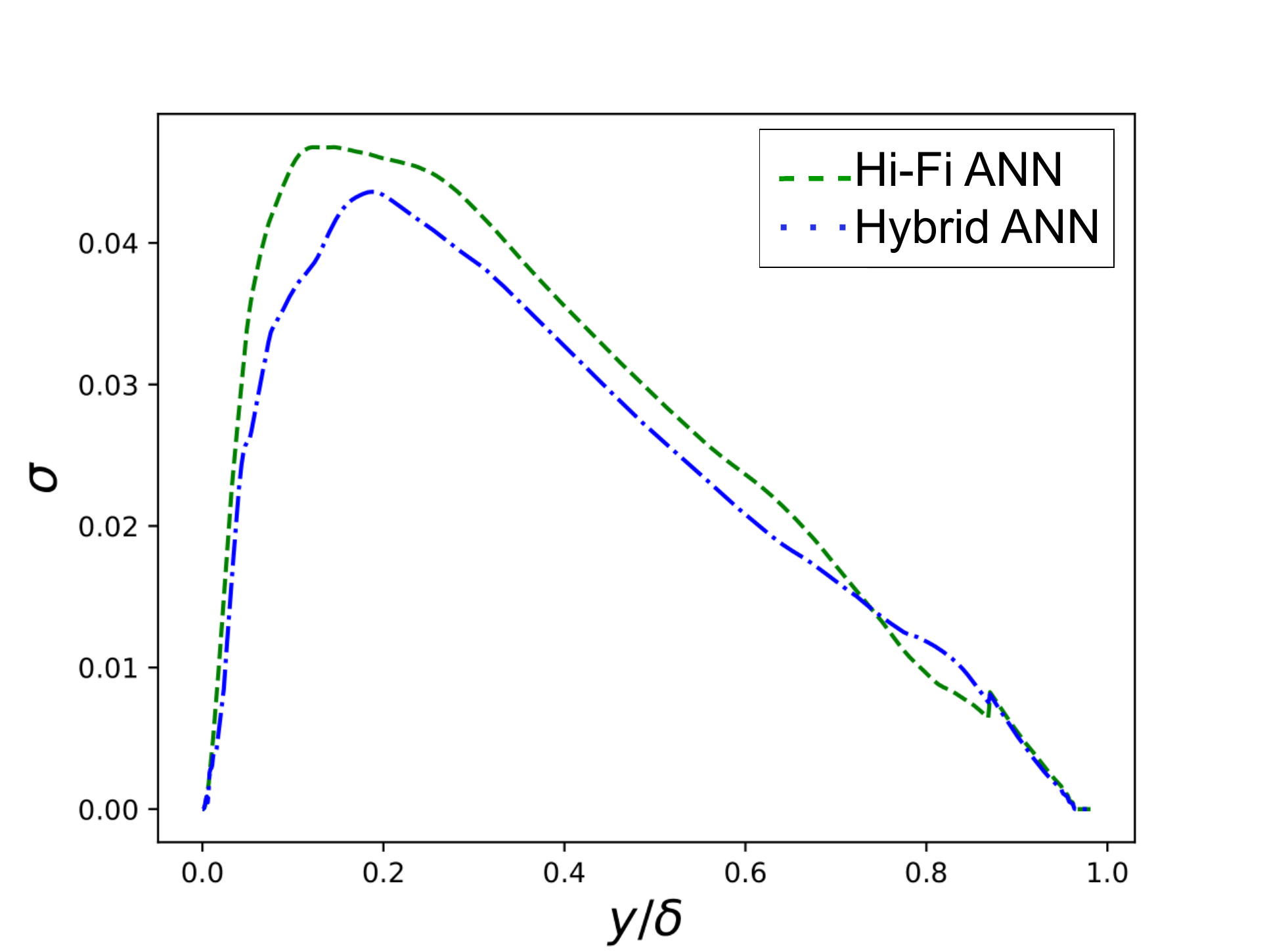}} 
	\caption{Distribution of the standard deviation of the streamwise and wall-normal heat flux components estimated by perturbing the true anisotropy turbulent state.}
	\label{sigma_momentum}
	
\end{figure*}

The standard deviation computed for the two models at both values of the Prandtl numbers is shown in Figure \ref{sigma_momentum}. The standard deviation of the high-fidelity model achieves its peak in the near wall region, where the dependence of the model on the anisotropy state is maximum. In this region, the standard deviation of the hybrid model predictions is much lower since the training with both data mitigates the dependency on the Reynolds stress anisotropy. This evidence agrees with the considerations drawn for the Shapley values for $y^+$ close to 10.0. This region of peak turbulence production, where temperature gradients are most significant, and the influence of the temperature field on heat flux is at its maximum, is critical. Therefore, reducing the uncertainty in model predictions in this area is of utmost importance. These considerations lead to the conclusion that the hybrid ANN is significantly more robust than the high-fidelity ANN when it comes to handling inaccuracies in the combined momentum turbulence model.

\subsection{Results of the simulation test case}
\begin{figure*}[htb!]
	\includegraphics[width = 0.9\linewidth]{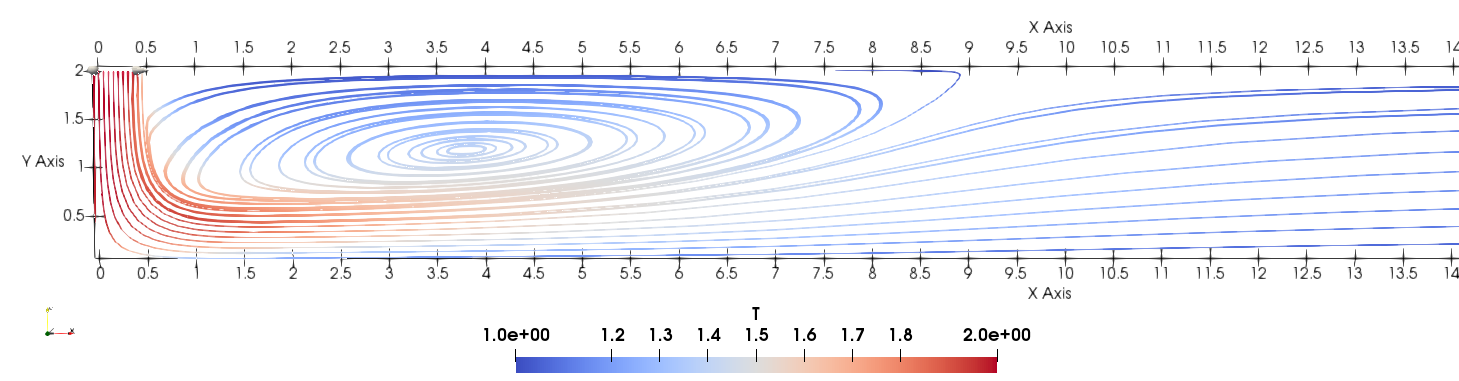}
	\includegraphics[width = 0.9\linewidth]{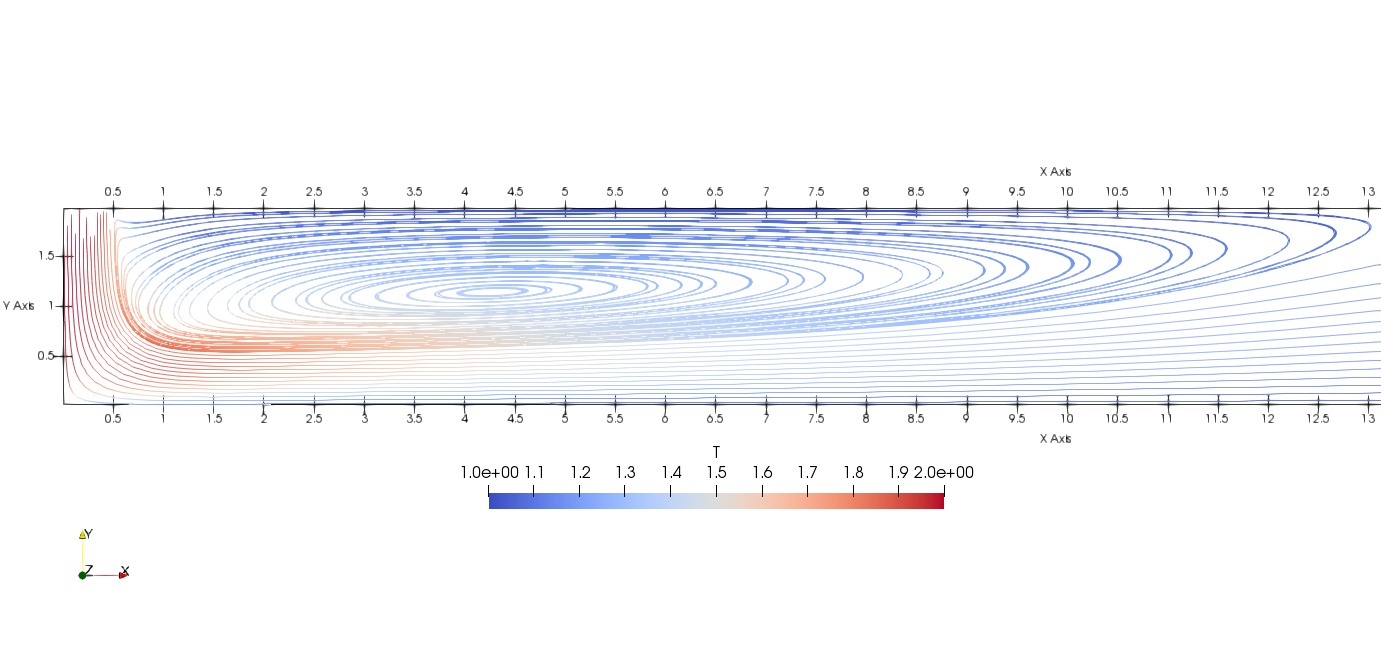}
		\caption{Shape of the recirculation bubble obtained with the Launder-Sharma $k-\epsilon$ model (top) and with the $k-\omega$ SST (bottom).}
	\label{velocity_imping_bubble}
\end{figure*}

\begin{figure*}[htb!]
    \centering
    \includegraphics[width = 0.8\linewidth]{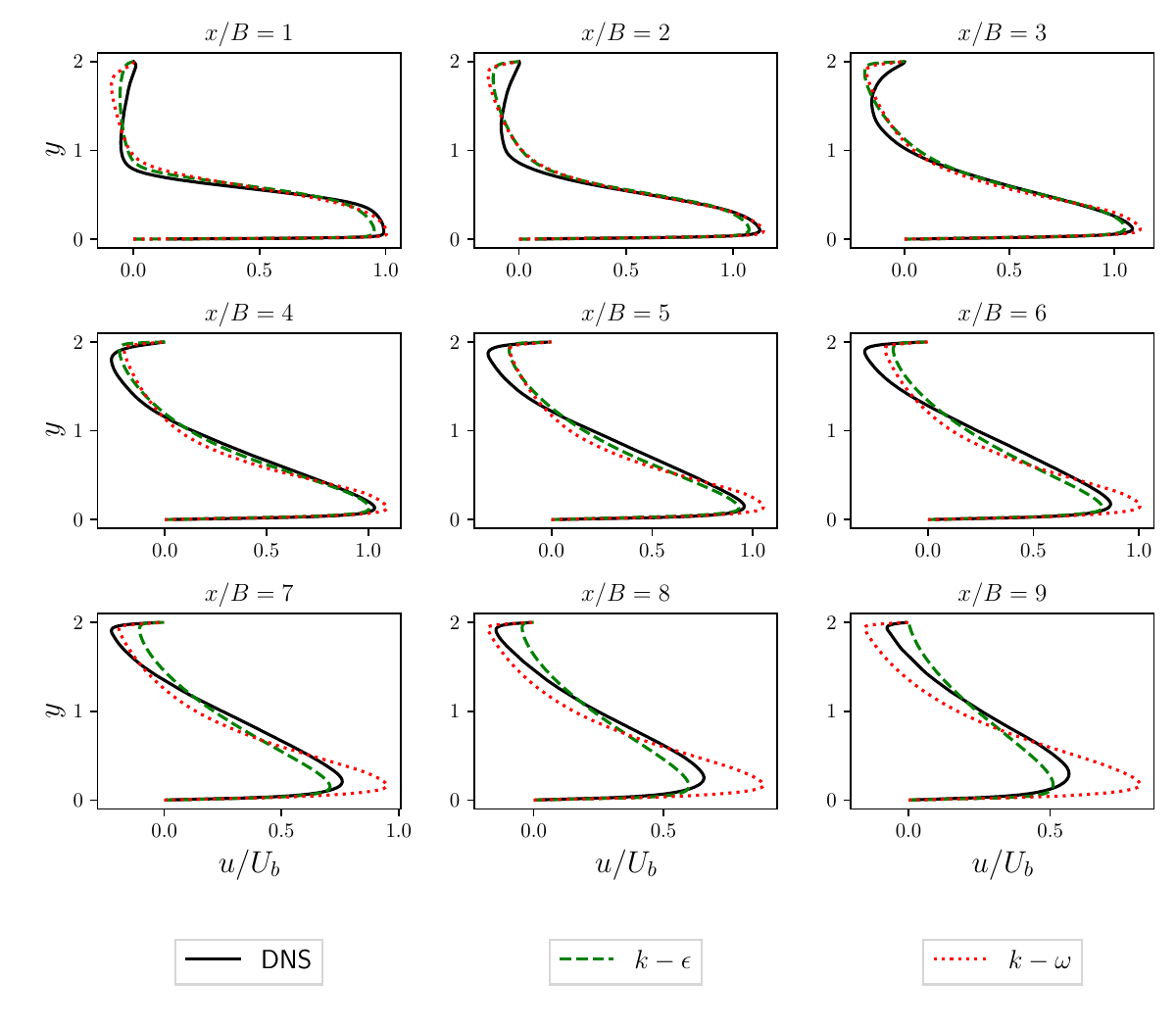}  
    \caption{Streamwise velocity profiles computed with different RANS momentum treatments at several distances from the impinging point ($x/B=0$). Comparison with the reference DNS data \cite{duponcheel2021direct}.}
    \label{vel_imping}
\end{figure*}

\begin{figure*}[htb!]
    \centering
     \includegraphics[width = 0.8\linewidth]{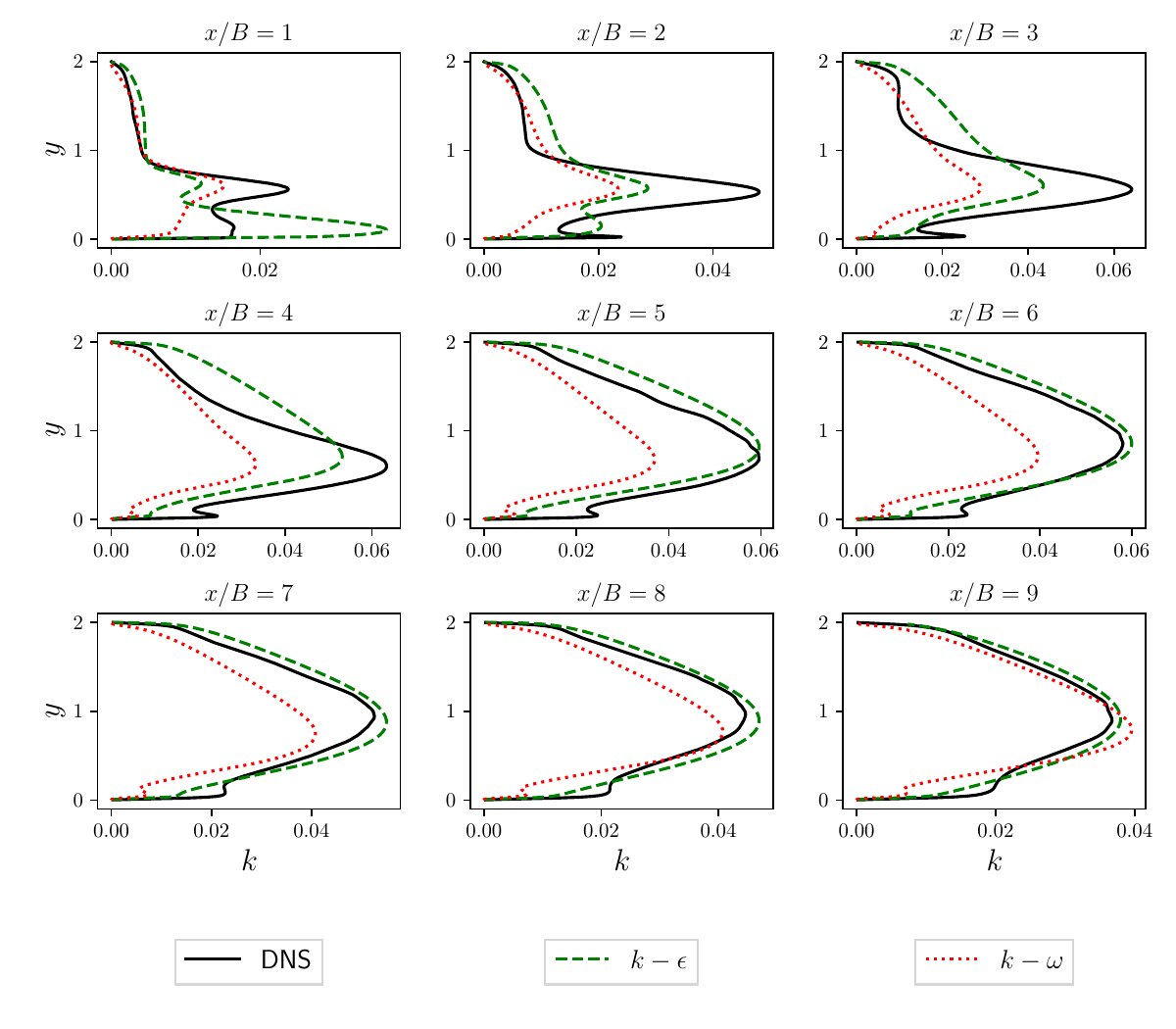}   
    \caption{Profiles of the turbulent kinetic energy computed with different RANS momentum treatments at several distances from the impinging point ($x/B=0$). Comparison with the reference DNS data \cite{duponcheel2021direct}.}
    \label{k_imping}
\end{figure*}

The numerical setup described in section \ref{CFD_sec} was employed to verify the behavior of the hybrid model when combined with LEVMs, and to compare the new formulation with the original data-driven AHFM. As anticipated in section \ref{CFD_sec}, for this test case, the computed velocity field highly depends on the type of momentum treatment and the choice of the specific LEVM. This is confirmed by Figure  \ref{velocity_imping_bubble}, which presents the velocity fields computed with the $k-\epsilon$ and $k-\omega$ models. The recirculation bubble's size and shape generated after the impinging point significantly differs from the two closure models. The difference can also be appreciated from the velocity and $k$ profiles reported in Figure \ref{vel_imping} and \ref{k_imping}, where they are compared with the reference DNS. In particular, the $k-\epsilon$ and $k-\omega$ models underestimate and overestimate the recirculation bubble's extension.  

The data-driven thermal model and the Manservisi model \cite{manservisi2014cfd} were applied over the two underlying momentum fields to compute the heat flux and temperature distributions. The high-fidelity ANN, whose predictions are indicated with dashed blue lines, completely mismatches the reference heat flux profiles due to the inconsistency between training (DNS) and simulation (RANS) momentum data. The hybrid ANN gives instead accurate heat flux predictions, regardless of the combined momentum closure ($k-\epsilon$ or $k-\omega$). These results show that the applicability of the hybrid ANN is not restricted to the momentum closure that generated the training data (Launder-Sharma $k-\epsilon$), but could potentially extend to the entire family of LEVMs. The heat flux fields given by the hybrid ANN are close to the one computed with the Manservisi model, although a slight improvement in the accuracy can be appreciated close to the slit. 

\begin{figure*}[h!]
	\centering
	\includegraphics[width = 0.6\linewidth]{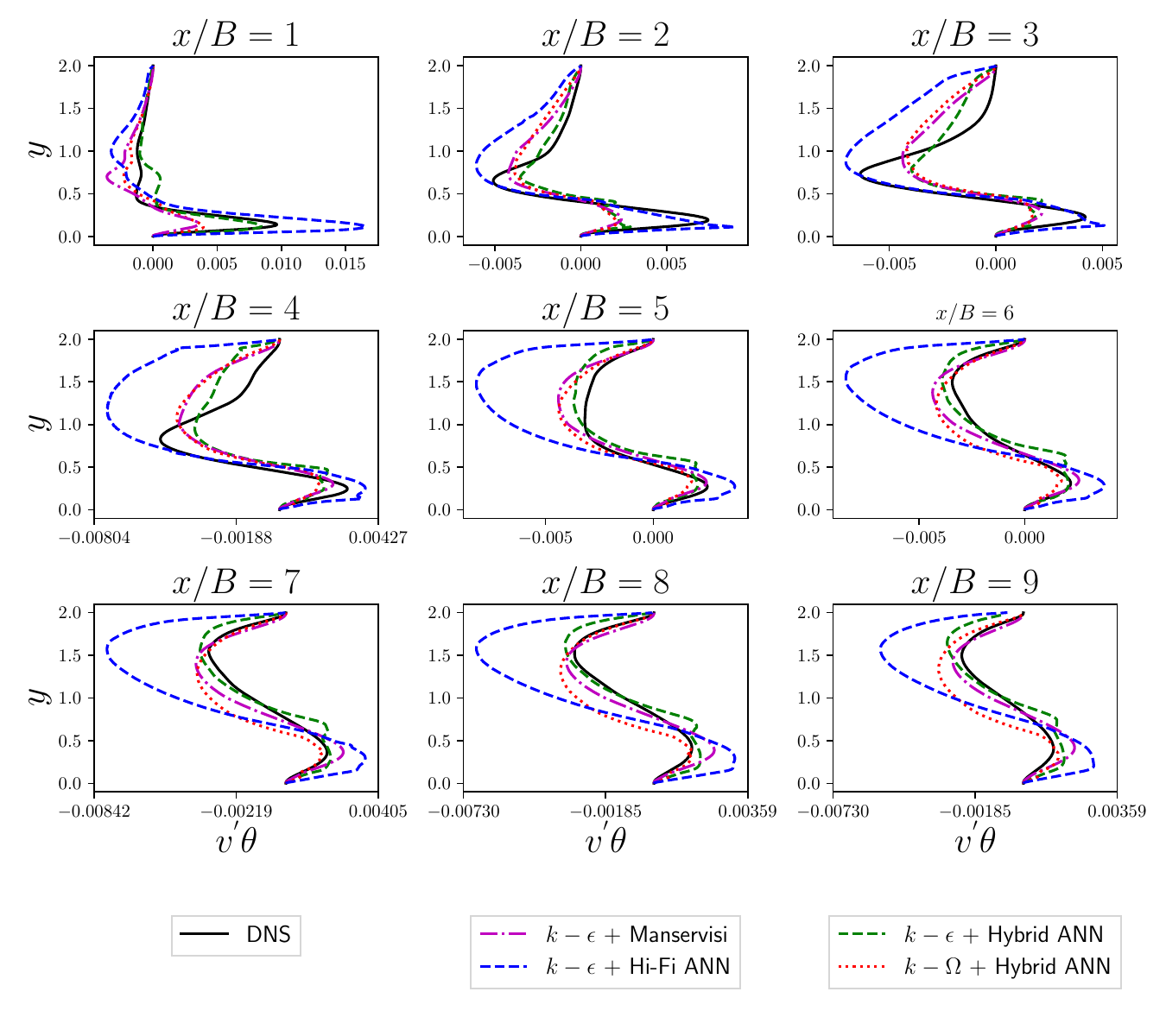}
		\includegraphics[width = 0.8\linewidth]{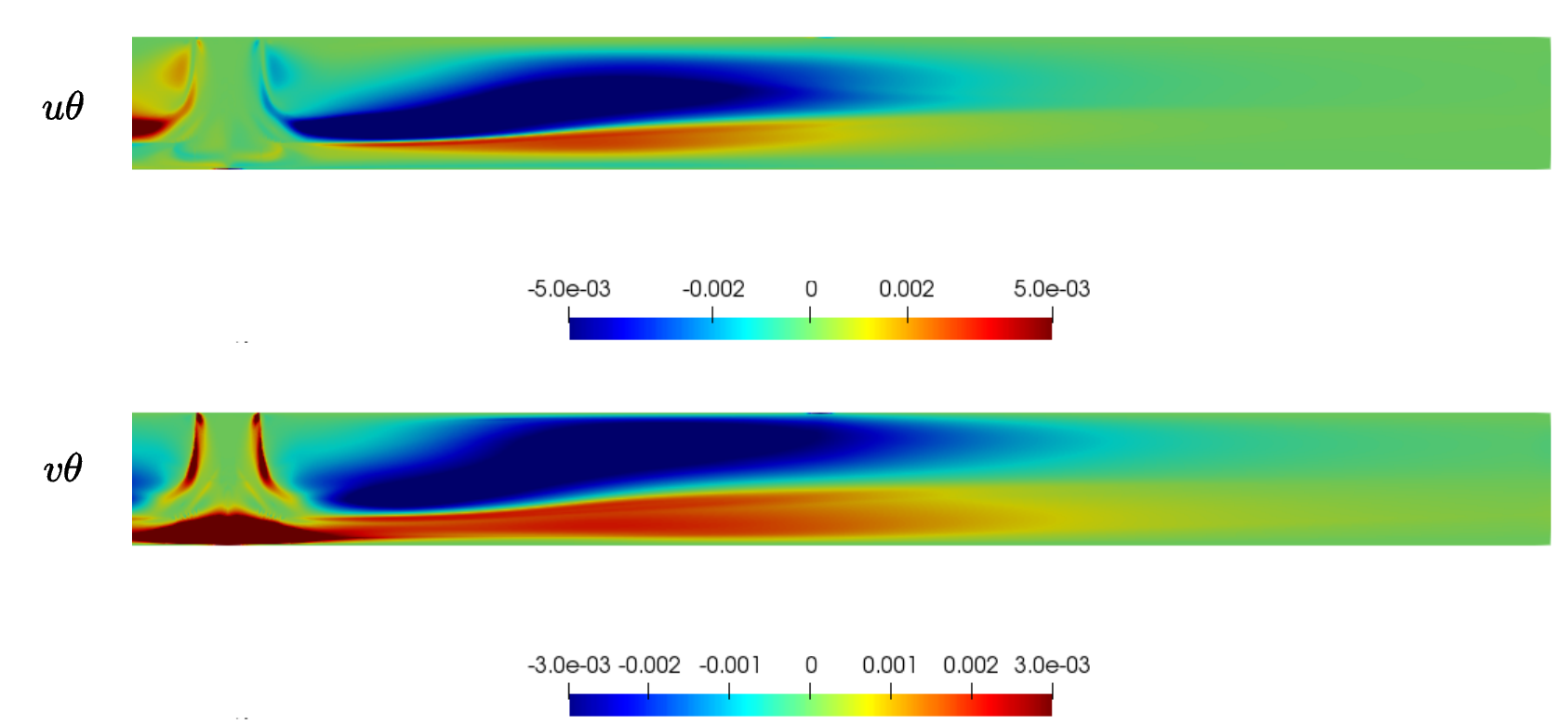}
	\caption{Top: wall normal heat flux profiles obtained with the hybrid ANN, the Hi-Fi ANN and Manservisi model \cite{manservisi2014cfd}. Comparison with the DNS data \cite{duponcheel2021direct} at several distances from the impinging point. Bottom: contours of the streamwise ($\overline{u \theta}$) and wall normal ($\overline{v \theta}$) heat fluxes.}
	\label{wn_heatflux_imping}
\end{figure*}

\begin{figure*}[h!]
	\centering
	\includegraphics[width = 0.6\linewidth]{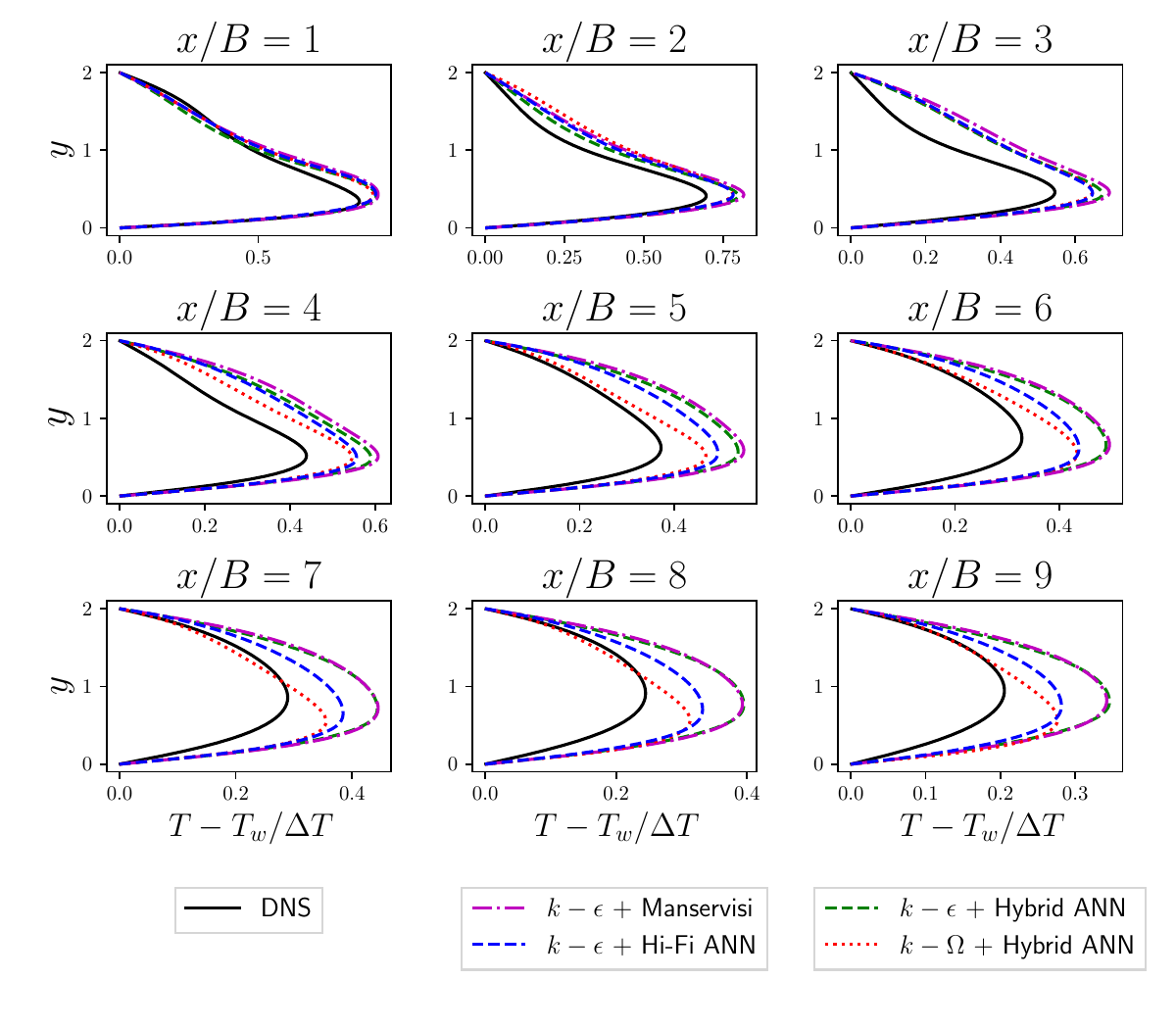}
					\includegraphics[width = 0.8\linewidth]{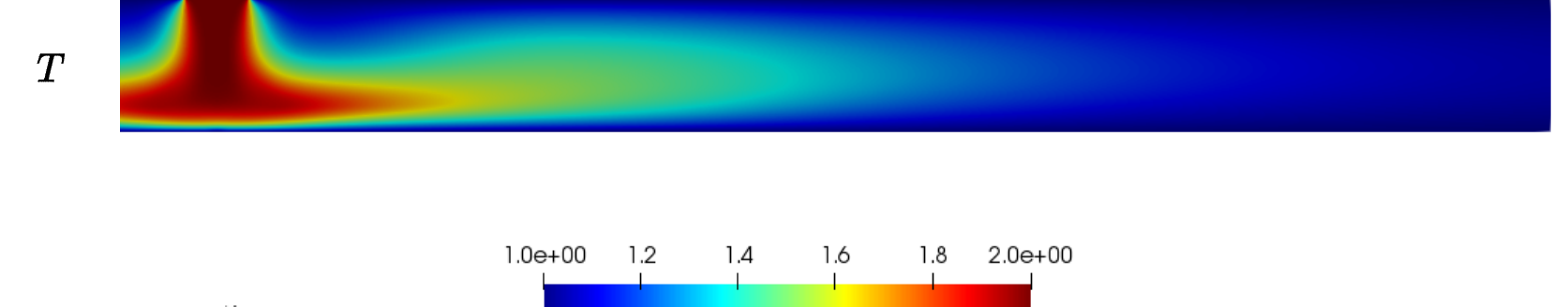}
	\caption{Top: temperature profiles obtained with the hybrid ANN, the Hi-Fi ANN and the Manservisi model \cite{manservisi2014cfd}. Comparison with the DNS data \cite{duponcheel2021direct} at several distances from the impinging point. Bottom: contour of the temperature field.}
	\label{temperature_imping}
\end{figure*}

The comparison among the thermal models in terms of temperature distribution is presented in Figure \ref{temperature_imping}, in which the hybrid ANN combined with the $k-\omega$ SST shows the best accuracy, while the hybrid ANN and the Manservisi model significantly overestimate the temperature, especially far from the slit. However, the better thermal field computed with the $k-\omega$ and hybrid ANN does not originate from a significantly more accurate heat flux modeling. Indeed, the comparison of the profiles in Figure \ref{wn_heatflux_imping} does not indicate the superiority of the heat flux predictions compared to the other two setups ($k-\epsilon$-hybrid ANN and $k-\epsilon$-Manservisi model). The better temperature agreement obtained with the high-fidelity ANN than with the hybrid ANN with the same momentum closure suggests that this temperature field is a result of compensation errors of both momentum and thermal closures. This consideration further underlines the importance of combining accurate thermal and momentum closures for heat transfer RANS simulations, as both the turbulent treatments contribute substantially to the resulting thermal field. This is especially true for low Prandtl numbers, at which the turbulent heat transport can be comparable, or even lower, than molecular transport and convection contributions in the overall energy balance. 

\section{Conclusions and outlook}\label{sec6}

This article investigated the problem of model-data inconsistency for a data-driven thermal turbulence model trained with high-fidelity data. Specifically, we explored ways to improve the robustness of the data-driven model to extend its coupling with low-fidelity models for Reynolds stresses (LEVMs), generally preferred in industrial contexts for their stability and computational cost. 

The crucial point was understanding whether the modeled (RANS) input statistics could be employed in training to inform the data-driven model about the inconsistencies between low and high-fidelity data. Detecting the quality of the momentum treatment could allow to adapt to different inputs or mitigate the sensitivities to the most critical ones. For such purpose, the work proposes the analysis of the multi-fidelity input space and a training strategy based on a hybrid dataset consisting of DNS and RANS input data. 

The work demonstrates that increasing the robustness of the thermal model with respect to momentum modeling is possible. The systematic deviation between high-fidelity and low-fidelity statistics can be utilized to identify the type of momentum modeling and adapt the output accordingly or to find alternative relationships of the input statistics. The multi-fidelity training mode generates a model less sensitive to Reynolds stress anisotropy than the model trained with only high-fidelity data, especially in regions characterized by high anisotropy. The network naturally adapts its sensitivity to low-fidelity inputs (RANS) and is thus more robust than the original. From a structural standpoint, the hybrid model appears to incorporate Reynolds stress anisotropy modeling through algebraic expansions of tensors dependent on velocity gradients.

The validation of the new data-driven model for a non-isothermal planar impinging jet shows that it is significantly more robust than the original high-fidelity network and provides accurate predictions with different eddy viscosity models. The uncertainty propagation analysis demonstrates, more generally, the greater robustness of the new model against perturbations of the Reynolds stresses in the barycentric plane.

While more test cases are needed to develop a truly general-purpose model, the promising results of this work open the door to new training strategies for data-driven turbulence closures that can leverage information from databases of varying sizes and fidelity levels. As demonstrated here, multi-fidelity inputs enhance model robustness and resilience to perturbations. Additionally, incorporating multi-fidelity targets can expand the training database to include conditions where high-fidelity data are challenging to obtain, such as high Reynolds or Grashof numbers or flows involving multiple physics or complex geometries. This approach can improve the generality and applicability of the learned closures. Future work will focus on extending the current framework to a broader range of data, both in inputs and outputs, to further advance data-driven turbulence modeling and its application in industrial contexts.


\section*{Nomenclature}\label{nom_sec}

{\footnotesize
\begin{minipage}[t]{0.5\textwidth}
\begin{tabular}{l l l}

$\mathbf{U}$ & [m/s] & Mean velocity \\
$T$	& [K] & Mean Temperature\\
$\mathbf{u}$ & [m/s] & Velocity fluctuation \\
$\theta$	& [K] & Thermal fluctuation\\
$k$ & [m$^2$/s$^2$]	& Turbulent kinetic energy\\
$\epsilon$ & [m$^2$/s$^3$] & Turbulent dissipation rate\\
$k_{\theta}$ & [K$^2$] & Thermal variance\\
$\epsilon_{\theta}$	& [K$^2$/s] & Thermal dissipation rate\\
$\overline{\mathbf{u}\mathbf{u}}$	& [m$^2$/s$^2$] & Reynolds stress \\

$ $ \\

\end{tabular}
\end{minipage}
\begin{minipage}[t]{0.5\textwidth}
\begin{tabular}{l l l}

$\overline{\mathbf{u}\theta}$	& [mK/s] &	Turbulent heat flux\\
$\nu$	& [m$^2$/s] & Molecular viscosity\\
$\alpha_l$	& [m$^2$/s] & Molecular diffusivity\\
$\delta$	& [m] & Half channel width\\
$\mathbf{I}$	& [-] & Identity tensor\\
$\mathbf{S}$	& [1/s] & Strain rate tensor\\
$\boldsymbol{\Omega}$	& [1/s] & Rotation tensor\\
$\mathbf{b}$	& [-] & Reynolds stress anisotropy tensor\\
$ $ \\

\end{tabular}
\end{minipage}

}

\begin{acknowledgments}
This work was supported by an F.R.S.-FNRS FRIA grant, and the authors gratefully acknowledge Prof. Iztok Tiselj and Dr. Mathieu Duponcheel for providing their datasets.
\end{acknowledgments}

\appendix

\section{RANS setup for the training flows} \label{setup_flows_rans}
The simulations of the RANS counterpart of the flows in the training database were conducted in the openfoam environment with the Launder-Sharma $k$-$\epsilon$ model. The choice of the computational domains follow the setups described in Ref. \cite{kawamura2000dns} and \cite{oder2019direct}. For both setups, the mesh is structured and consistent with the wall-resolved approach, i.e., the values of $y^+$ range between 0.1 and 1.0. Table \ref{channel} and \ref{2d_rans_bfs} indicate the boundary conditions applied for the two flows. 

For the channel, the fully developed flow is obtained by imposing cyclic boundary conditions at the inlet and outlet and a pressure gradient source term that adapts to the bulk velocity prescribed. For the backward-facing step flow, a fully developed flow in the channel preceding the step is obtained using recycling conditions, i.e., the iterative remapping of the fields at a certain distance from the inlet to the inlet boundary.

\begin{table*}[h!]
	\caption{Overview of the boundary conditions imposed for the RANS simulation of the backward facing step.}
	\label{2d_rans_bfs}
	\centering
		\begin{tabular}{lllll}
			\hline
			Field &
			Inlet &
			Outlet &
			Walls \\ \hline
			$\mathbf{U}$ &
			Recycling &
			$\frac{\partial \mathbf{U}}{\partial n}=0$ or $\mathbf{U} \cdot n =0$ &
			$\mathbf{U}=0$ \\
			$p$ &
			$\frac{\partial p}{\partial n }=0$ &
			Fixed: $p=0$ &
			$\frac{\partial p}{\partial n }=0$ \\
			$k$ &
			Recycling &
			$\frac{\partial k}{\partial n}=0$ &
			$k=0$ \\
			$\epsilon$ &
			Recycling &
			$\frac{\partial \epsilon}{\partial n}=0$ &
			$\epsilon=0$ \\
			$\mathbf{\overline{uu}}$ &
			Recycling &
			$\frac{\partial \mathbf{\overline{uu}}}{\partial n}=0$ &
			$\mathbf{\overline{uu}}=0$  \\ \hline
		\end{tabular}%
	
\end{table*}

\begin{table*}[h!]
	\caption{Overview of the boundary conditions imposed for the RANS simulation of turbulent channel flows.}
	\label{channel}
	\centering
		\begin{tabular}{lllll}
			\hline
			Field &
			Inlet &
			Outlet &
			Walls \\ \hline
			$\mathbf{U}$ &
			Cyclic &
			Cyclic &
			$\mathbf{U}=0$ \\
			$p$ &
						Cyclic &
			Cyclic &
			$\frac{\partial p}{\partial n }=0$ \\
			$k$ &
			Cyclic &
			Cyclic &
			$k=0$ \\
			$\epsilon$ &
			Cyclic &
			Cyclic &
			$\epsilon=0$ \\
			$\mathbf{\overline{uu}}$ &
			Cyclic &
			Cyclic &
			$\mathbf{\overline{uu}}=0$  \\ \hline
		\end{tabular}%
	
\end{table*}

\bibliography{references}

\end{document}